\documentstyle[prb,aps,twocolumn]{revtex}
\include{psfig}

\title{ Dissipative Electron Transport through Andreev Interferometers
}  
\author{
	 H. A. Blom$^{1}$, A. Kadigrobov$^{1,2}$, A. M. Zagoskin$^{3}$, R. I. Shekhter$^{1}$, and M.~Jonson$^{1}$ }

\address{
 $^{1}$Department of Applied Physics,
 Chalmers University of Technology and G\"oteborg University, 
SE-412 96 G\"oteborg, Sweden\\
 $^{2}$B. I.  Verkin Institute for Low Temperature Physics \& 
Engineering, \\
National Academy of Science of Ukraine, 4 Lenin Ave., 310164 Kharkov, Ukraine\\
 $^{3}$Physics and Astronomy Dept., Univ. of British Columbia, 6224 Agricultural Rd., B.C., Canada V6T 1Z1
 }

\begin{document}

\date{G{\"o}teborg preprint APR 97-5, 24 June 1997}
\maketitle

\begin{abstract}
We consider the conductance of an Andreev interferometer, i.e., a 
hybrid structure where
a dissipative current flows through a mesoscopic normal (N) sample 
in contact with two
superconducting (S) ``mirrors". Giant conductance oscillations are 
predicted if the 
superconducting phase difference $\phi$ is varied. Conductance maxima 
appear when $\phi$
is on odd multiple of $\pi$ due to a bunching at the Fermi energy of 
quasiparticle energy levels 
formed by Andreev reflections at the N-S boundaries. For a ballistic 
normal sample the oscillation
amplitude is giant and proportional to the number of open transverse 
modes. 
We estimate using both analytical and numerical methods how scattering 
and mode mixing --- which
tend to lift the level degeneracy at the Fermi energy --- effect the giant 
oscillations. These are 
shown to survive in a diffusive sample at temperatures much smaller than
the Thouless temperature provided there are potential barriers between 
the sample and the 
normal electron reservoirs. Our results are in good agreement with 
previous work on 
conductance oscillations of diffusive samples, which we propose can 
be understood in terms of 
a Feynman path integral description of quasiparticle trajectories. 

\end{abstract} 
	\section{Introduction}\label{introduction}
Recently considerable attention has been devoted to mesoscopic
superconductivity, i.e. to the
transport properties of mesoscopic systems with mixed normal (N) and
superconducting (S) elements, where new quantum interference effects have been
discovered. Novel physics appear 
in such systems because electrons undergo
Andreev reflections \cite{Andreev} at the N-S boundaries, whereby the 
macroscopic phase of a superconducting condensate is imposed on the 
quasi-particle wavefunctions in the normal regions. If transport in the
normally conducting part of the sample is phase coherent, there
is a possibility that interference between Andreev scattering at two (or more)
N-S interfaces makes the conductance of the hybrid system sensitive to the
phase difference $\phi$ between the superconducting elements; in this case one
may describe the system as an Andreev interferometer.

\begin{figure}
\vspace{.5cm}
%%%[hbt]\hskip 1cm \hbox
\centerline{\psfig{figure=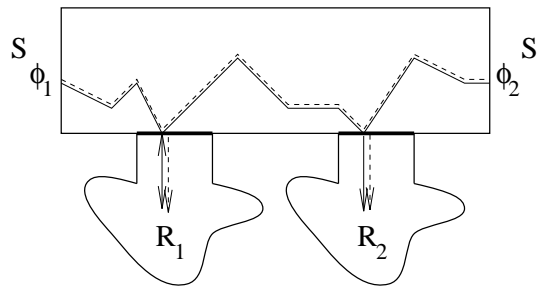,width=7 cm}}
\vspace*{0.5cm}
\caption{\protect{\label{fig22}}
Schematic picture of an Andreev interferometer consisting of
a normal (N) metal (diffusive transport regime) in contact with 
two superconducting
elements (S), which are characterized by the phases $\phi_1$ and $\phi_2$ of
their respective order parameter. The normal metal are in contact with two
reservoirs R$_1$ and R$_2$ via leads. The thick lines at the junctions between
the leads and the normal metal represent potential barriers, which act as beam
splitters partially reflecting quasiparticles 
impinging on the junctions, cf. Fig.~2. 
If transport is phase coherent
quasiparticles at the Fermi level (zero excitation energy) are phase conjugated
after Andreev reflection at the N-S interfaces, so that Andreev 
reflected holes 
(dashed line) retrace the path of the incoming electrons (full line) 
and vice versa.
}
%%%\vskip .5cm
\end{figure}

This paper is concerned with a theoretical description of hybrid mesoscopic 
systems of the Andreev
interferometer type. In particular we are interested in the normal conductance
as a function of the phase difference between the condensates of two separate
superconducting elements acting as ``mirrors'' by reflecting
 the qausi-particles
in the normally conducting element, which in its turn is 
connected to two electron reservoirs
 as schematically shown in Fig.~\ref{fig22}. 
The normal electron transport may be in the 
ballistic regime or in the diffusive regime; both cases will be discussed. 
In addition we will make the 
important distinction between the cases when potential barriers or (sharp)
geometrical features serve
as ``beam splitters'' at 
the junctions between the leads and the normal element and when the passage
between leads and sample is unhindered by quantum-mechanical
scattering between distinct quasi-classical trajectories at these junctions. 

The rest of this introduction will be divided into two parts: (i) a general
introduction to the subject of Andreev interferometry and (ii) a  qualitative
discussion based on quasi-particle trajectories which makes it possible to
understand the main features of the conductance oscillations of various
types of Andreev interferometers a a function of the superconducting
phase difference.

\subsection{Origin of conductance oscillations in Andreev interferometers}

Already in the early 80's
Spivak and Khmel`nitskii showed \cite{Spivak} the weak localization 
corrections to the conductance of a diffusive sample containing two 
superconducting mirrors to be sensitive to the superconducting 
phase difference. The effect can be understood in terms of the usual 
interpretation of weak localization as due to coherent backscattering.
The interference of probability amplitudes for classical quasi-particle
trajectories (or ``Feynman paths'') bouncing off both mirrors will depend
on the phases of the respective condensates. 
Considering a closed diffusive path touching both S-N interfaces --- where
electrons will be reflected as holes and vice versa --- the 
interference between quasi-particles moving in opposite directions, 
clockwise and 
counterclockwise around the path, results in a phase difference of
2$\phi$ between the interfering amplitudes, i.e. twice the phase difference 
between the two superconductors.
This is because the phase picked up due to Andreev reflections off the 
two mirrors is $\pm\phi$ depending on whether the motion is clockwise or
anticlockwise.
%
%As the superconducting phase difference $\phi$ gained by the particles 
%at the N-S boundaries has the opposite signs for particles moving clockwise 
%and counterclockwise along the path, the interference results in the 
%phase difference of the interfering amplitudes 
%with a loop  touching  superconductors,  
%where the electron-hole and the hole-electron 
%transformations take place due to Andreev reflections. 
%
It follows that the weak localization correction to the conductance
of a normal sample with two superconducting mirrors has a component
that oscillates with a period equal to $\pi$ as the 
phase difference between the superconductors is varied. 

%If the length $L$ of the normal conductor is less than 
%the phase breaking length $L_{\phi}$,  the value of these weak localization 
%corrections is of the order of $\hbar /p_F l_0 \ll 1$  ($p_F$ is the Fermi 
%momentum, $l_0$ is the mean free path length) and does not depend on the 
%ratio between $L$ and the coherence length of the normal metal,  
%$L_T=\sqrt{\hbar D/k_B T}$ ($D$ is the diffusivity, $k_B$ is the Boltzmann 
%constant, $T$ is the temperature).

In the beginning of the 90's, a  dependence on the phase difference $\phi$ 
was discovered not only for the conductance fluctuations but for the main 
conductance as well. Not only conductance fluctuations but the 
ensemble-averaged 
conductance itself can therefore be controlled by the phase difference 
between two  
superconducting mirrors. \cite{Takayan,Petr1,Hekk Naz,Lamb1,Lambert1}
Hybrid S-N systems (Andreev interferometers) 
which show such a behaviour at very low temperatures have 
lead-sample junctions which act as ``beam splitters'' in 
the sense that a quasi-particle approaching the junction along a
quasi-classical trajectory is only
partially transmitted. Hence, 
%
%as illustrated in Fig.~\ref{fig11} --- where
%we imagine potential barriers to act as beam splitters -- 
%
a beam splitting junction has the effect of
partly reflecting a quasi-particle coming from one N-S interface towards
the second N-S interface, as illustrated in Fig.~\ref{fig22}. This has the
important consequence that when a quasi-particle finally leaves the sample
to contribute to the current there is a certain probability for it to
have interacted with {\em both} superconducting 
mirrors. To be specific, an electron entering the sample from one reservoir, 
$R_1$ say (referring to Fig.~\ref{fig22}),
may 
follow a trajectory (full line) 
where first it is reflected as a hole by one mirror, 
say $S_{\phi_1}$,
then it returns (dashed line in Fig.~\ref{fig22}) to
bounce off the same junction through which it entered, gets reflected
towards the second superconducting mirror $S_{\phi_2}$ where it
is Andreev reflected
as an electron, and finally it passes (full line) through the junction 
to the second reservoir $R_2$ now carrying information about the 
difference $\phi=\phi_1 -\phi_2$ between
the phases of the two superconducting mirrors (the difference appears
because the phase picked up on Andreev 
reflection differs in sign between an electron 
and a hole). It follows \cite{Takayan,Hekk Naz}
 that such trajectories contribute a term to the
conductance that oscillates with period 2$\pi$ (rather than $\pi$)
as function of $\phi$.
%
%there is a beam splitter of a normal metal 
%(see Fig.~\ref{fig11}) that divides an electron wave incident from a 
%normal electron reservoir, between the two other branches coupled to 
%superconductors. This causes the spatial interference of the electronic 
%waves after electron-hole and hole-electron transformations at the N-S 
%boundaries resulting in $2\pi$-periodicity of the conductance 
%\cite{Takayan,Hekk Naz}. 

\begin{figure}%%%[hbt]\hskip 1cm \hbox
\centerline{\psfig{figure=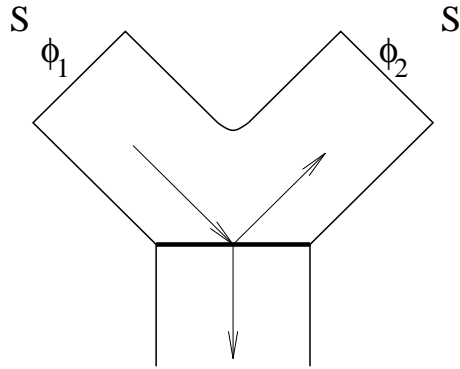,width={6 cm}}}
\vspace{.5cm}
\caption{\protect{\label{fig11}}
 A potential barrier at the junction between the normal metal and
the lead to a reservoir splits the quasiparticle beam 
coming from one of the superconducting mirrors (cf. arrows). This makes it
possible for quasiparticles having undergone Andreev reflection at both mirrors
to contribute to the current even if their excitation energy is zero and therefore
reflected hole (electron) excitations retrace the paths of the electron (hole)
excitations. It follows that the conductance may depend 
on $\phi_1 - \phi_2$  even at zero temperature (see text).
}
%%%\label{fig11}}\vskip .5cm
\end{figure}

As we have indicated above the influence of the superconducting phase 
difference on the conductance 
of an Andreev interferometer structure is an interference effect.  
The macroscopic phases of the superconducting condensates, or --- using 
a different language --- of the order parameter or of the gap function of 
the respective superconducting mirror are imposed on the microscopic
wavefunctions of the electron- and holelike quasi-particles when they
undergo Andreev scattering at the N-S boundaries. The dominating role of 
these scattering phases is due to the effect of compensation of the phases 
gained along the electron- and hole sections of the trajectories connecting 
the Andreev reflection. Returning to Fig.~\ref{fig22}, we note that it 
illustrates how an electron (hole) with energy infinitely close to but
above (below) the 
Fermi energy follows a trajectory [full (dashed) line] towards an
N-S interface. When it gets Andreev reflected as a hole, conservation of
energy and momentum makes its energy infinitely close to but below (above)
the Fermi energy and the hole (electron) retraces the path 
[dashed (full) line] of the incoming
electron (hole). In this way the phase acquired by an electron is 
``eaten up'' as the hole retraces the electron path in the opposite direction
and the net change of phase is due to the Andreev reflection only.

The possibility of phase compensation
only exists  for quasi-particles whose energies are very close to the Fermi 
energy.
Because energy and momentum is conserved in the Andreev
scattering process a quasi-particle with excitation energy $E$ measured
from the Fermi energy $\epsilon_F$ is reflected as a hole of energy $-E$
and in a direction that 
differs from the incoming path by an angle of order $E/\epsilon_F$.
This implies that for finite quasi-particle excitation energies the
phase compensation will not be complete. Since the dominating role of
the superconducting phase difference is lost when the uncompensated 
phase along the quasi-particle trajectory connecting the two superconducting
mirrors is of order $2\pi$ it follows immediately that only quasi-particles
whose excitation energies are less than a critical energy $E_{c}$
may contribute to the $\phi$-dependent part of the conductance.
For ballistic samples $E_{c}\sim \hbar v_F/L$, while 
$E_{c}\sim \hbar D/L^2$ in the diffusive transport regime, where it is known
as the Thouless energy
($v_F$ is the Fermi velocity, $D$ is the diffusion constant).

The restriction on quasi-particle excitation energies translates into
a temperature dependence, where the Thouless energy sets the
characteristic temperature scale. Nazarov {\em et al.}
\cite{Nazarov Stoof} and Volkov {\em et al.} \cite{Volkov Lambert}
have for example suggested
a ``thermal mechanism'' that gives a large amplitude of the $2\pi$-periodic
conductance oscillations with $\phi$
at temperatures close to the Thouless temperature. Their result is
 due to a dependence of the 
effective diffusion coefficient on the energy of the quasiparticles in a 
hybrid S-N-S sample and will be further discussed below.

In addition to dephasing effects due to finite excitation energies 
phase coherence may be broken by inelastic scattering.
%
% completely controls the interference 
%effect if only the non-compensated  phase picked  along the trajectories 
%between two Andreev reflections is not larger than $2\pi$. Violation of 
%this condition results in suppresion of the phase tuning of the conductance. 
%This suppression occurs if  energies of the quasiparticles are greater 
%than the Thouless energy $E_{Th}$ ($E_{Th}\sim \hbar v_F/L$ for ballistic 
%samples, and $E_{Th}\sim \hbar D/L^2$ for diffusive samples).
%
The interference effects described can therefore only be observed if 
the length $L$ of the normally conducting part of the sample is at most
of the order of the phase breaking length $L_{\phi}$
or the normal metal coherence length $L_T$, whichever of them is smaller.
In the ballistic
transport regime $L_T =\hbar v_F/k_B T$, while in the diffusive transport 
regime one has $L_T=\left(\hbar D/k_B T\right)^{1/2}$ 

A large number of experimental and theoretical investigations
%\cite{Poth1}-\cite{Petrashov3}.
\cite{Nazarov Stoof,Volkov Lambert,Poth1,Wees2,Vegvar,Naz,Zaitsev,Beenakker,Petrashov,Claughton95a,go,Lambert,Claughton,Allsopp,Claughton Lambert,Claughton96b,Volkov Zaitsev,Mur96,Charlat96,Hartog96,Rahman,Ohki96,Weesbook,Spivak Zhou,Petrashov3} followed the early work on the tunable conductance of mesoscopic 
samples of the Andreev interferometer type. 
For  diffusive samples 
the amplitude of the conductance oscillations has been found  to 
be large in the sense that it is comparable to
 the conductance in the absence of  superconducting 
elements. The conductance maxima usually appear 
at even multiples of $\pi$. As discussed by Kadigrobov {\em et al.}
\cite{go} the situation is quite different for ballistic Andreev 
interferometers, where the conductance oscillations may be giant ---
i.e. the oscillation amplitude may be much larger than the conductance in
absence of superconducting mirrors.
The system discussed in Ref.~\onlinecite{go} is shown in Fig.~\ref{fig22};
the normal part of a hybrid S-N-S system is weakly coupled to two
normal electron reservoirs and hence the dissipative current flows from 
one reservoir to the other via the normal metal element. Two 
low-transparency barriers form the junctions
between sample and leads (going to the reservoirs) and act as beam splitters
in the sense outlined above.

%
%In paper \cite{go} giant conductance oscillations  were predicted for 
%a hybrid S-N-S ballistic system where the normal part  was  weakly 
%coupled to two normal electron reservoirs (that is there were  two  
%barriers of low transparancy  between the sample and the leads to the 
%reservoirs, see Fig.~\ref{fig22}), and therefore the dissipative current 
%flew from one reservoir to the other via the normal part of the system.
%The amplitude of the predicted conductance oscillations can greatly exceed 
%the  conductance in the absence of superconducting elements, being equal 
%to the maximal possible conductance 
%$G= N_{\perp} e^2/h$ ($N_{\perp}$-the number of the open transverse modes). 
%The period of  oscillations is also $2\pi$ but the resonant peaks of the 
%conductance
%are at odd multiples of $\pi$.

The giant conductance oscillations arises because the structure considered
in Ref.~\onlinecite{go} permits resonant transmission of electrons and holes 
via  the normal part of the sample. Resonant transmission
occurs when the spatial quantization of the electron-hole motion in the 
mesoscopic normal element leads to allowed energy levels coinciding with  
the Fermi energy (at zero temperature and small bias voltage the energy of the 
electrons incident from the source reservoir is equal to the Fermi energy). 
It follows from the semiclassical Bohr-Sommerfeld quantization rule 
[cf. Eq.~(\ref{Aspectr}) below]
that all the $N_{\perp}$ conducting transverse modes 
in the normally conducting element have one quantized level 
at the Fermi energy if the phase 
difference $\phi$ between the two superconductors is equal to an odd 
multiple of $\pi$. This means that for 
$\phi=\pi(2k+1),\:k=0,\pm 1, \pm 2...$ 
each transverse mode  can resonantly transmit  electrons, and 
hence $N_{\perp}$ transverse modes contribute to the resonance 
simultaneously. As a result the amplitude of the conductance oscillations 
reaches the maximal value $G_{max}=N_{\perp}2e^2/h$ when $\phi$ is an odd 
multiple of $\pi$  (giant oscillations).

\subsection{Understanding conductance oscillations in Andreev 
interferometers in terms of Feynman paths}

In all experimental and theoretical studies of Andreev interferometers
three types of quasi-particle scattering mechanisms (in various combinations) 
have to be taken into account. Scattering of charge carriers can be due to:

\begin{enumerate}
\item
potential barriers or geometrical feautures (beam splitters) at the 
junctions between the mesoscopic 
sample and the leads to the electron reservoirs

\item 
impurity scattering inside the mesoscopic region 

\item
non-Andreev (normal) reflection from potential barriers at the N-S interfaces.

\end{enumerate}

Here we shall emphasize the crucial role played by
beam splitters in distinguishing between different types of oscillation 
phenomena. Therefore we choose to separately discuss two different types  of 
hybrid S-N-S structures: those with- and those without beam splitters. In
particular we will show below that 
the presence of beam splitters is necessary for conductance
oscillations with $\phi$ to appear in the limit of vanishing temperature.

\subsubsection{Andreev interferometers without beam splitters}

% (thermal interference effect for the conductance).

\begin{figure}
%%%[hbt]\hskip 1cm \hbox
\centerline{\psfig{figure=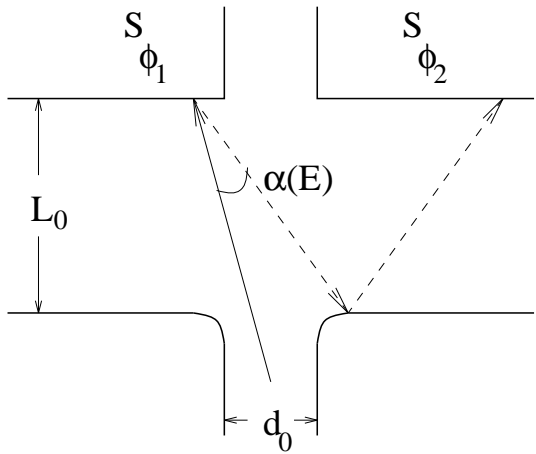,width=7 cm}}
\vspace{0.5cm}
\caption{\protect{\label{fig33}}
At finite excitation energies $E$ the path of an Andreev reflected 
hole (electron) in a ballistic system deviates by a finite angle $\alpha(E)$ 
from the path of the incoming electron (hole). If $E$ is sufficiently large 
--- but not otherwise --- one quasiparticle trajectory may therefore,
as shown here, hit both 
superconductors.
}
%%%\label{fig33}}\vskip .5cm
\end{figure}

In the absence of beam splitters
quasiparticles are ballistically injected into the mesoscopic sample along
quasi-classical trajectories
without suffering any quantum-mechanical scattering between trajectories
at the junctions between sample and 
(leads going to the) reservoirs. 
In this case the quasi-particles therefore freely pass the contact region 
without undergoing reflection. It is not difficult to convince oneself
that in such a system there 
are no low energy quasi-particle trajectories connecting the reservoir 
(or reservoirs) and {\em both} superconductors.  This is because 
a quasi-particle with vanishingly small excitation energy
is perfectly
backscattered at the N-S boundaries in the sense that the angle of Andreev
scattering is equal to $\pi$.
Therefore such a trajectory can not connect more than two 
bodies (say, the reservoir and one of the N-S interfaces). Of course, if the 
energy increases, then due to inelastic Andreev reflection (an electron 
with energy $E$ is transformed into a hole with energy $-E$; the total energy
is conserved since a Cooper pair of zero energy is created) the 
back-scattering is not perfect and, in contrast to the case when $E=0$, 
the angle of reflection differs from $\pi$ by a value 
$\alpha\approx E/\epsilon_F$. An interference effect involving the condensate
phases of both mirrors is now possible 
since an electron Andreev-reflected as a hole 
at the S-N interface follows a different trajectory 
than the impinging electron and hence
has a finite probability not to reach the injector region.
In this case, as shown in Fig.~\ref{fig33}, it is possible that the 
trajectory will reach the second superconductor before finally escaping to
a reservoir. One may readily evaluate  
the role of the described inelastic Andreev reflection in the formation of 
phase sensitive trajectories for both the ballistic and the diffusive case.

In the ballistic case an injected electron which is 
Andreev reflected as a hole will not 
directly return to the injector if the distance $L_0$ to the
superconducting mirror is large compared to the injector opening $d_0$.
The precise criterion is that it will not return if $\alpha L_0>d_0$,
where $\alpha\approx E/\epsilon_F$ 
(see Fig.~\ref{fig33}). If one takes into account that the excitation energy
as explained above is limited by the (ballistic) Thouless energy 
$E_c$ in order for phase 
coherence to be maintained, one concludes that an interference effect involving
both Andreev mirrors is possible only if the injector opening is smaller than
the electron deBroglie wavelength; this is because since
$\alpha(E)<\alpha(E_{c}=\hbar v_F/L_0)$ it follows that $\alpha L_0 <d_0$
if $d_0<\lambda_B$. (For a degenerate electron gas the deBroglie wavelength 
is equal to the Fermi wavelength.)
The inevitable conclusion is that an interference mechanism involving
thermally excited quasi-particles cannot play a role in realistic 
experiments using ballistic samples. Under these circumstances the effect
of scattering by impurities inside the mesoscopic sample 
is decisive for the desired interference phenomenon involving two
superconducting mirrors to occur. In other words --- in the absence of beam
splitters ---
we need to consider a mesoscopic sample in the diffusive transport regime.

In the diffusive case interference between Andreev scattering at two
spatially separated superconducting mirrors may occur 
if the  mirror-reflected trajectory diverges from the incident trajectory
by more than a de Broglie wavelength, $\lambda_B$, 
which we take to be the width of any particular trajectory.
In this case we may say that by inelastic Andreev reflection the reflected 
quasiparticle is sent into a different, classically distinguishable 
trajectory. When the 
separation becomes greater than $\lambda_B$ this trajectory interacts with 
a different set of impurities which will take the reflected particle on
a diffusive random walk along a completely different Feynman path. As 
the distribution of trajectories is homogeneous in the diffusive limit,  
there is 
a finite probability for the  trajectory (which starts from a reservoir)  to 
include points with Andreev reflections from both superconductors. This
implies that the criterion for the incident and
reflected trajectories to be sufficiently separated after a 
diffusing length of $L_D$ is $\alpha L_{D}\geq \lambda_B$. Since the angle
$\alpha \sim E/\epsilon_F$ this can be converted to a criterion for the
excitation energy of the form $E\geq E_{c}$. 
We recall that interference is destroyed for energies $E\gg E_{c}$.
Hence we conclude that there 
is a distinct group of quasiparticles with excitation energy 
around the Thouless energy $\sim E_{c}$ for 
which there is an interference effect controlled by the superconducting phase 
difference $\phi$.  As a result the temperature dependence of the conductance 
oscillations is non-monotonic. The amplitude of the oscillations vanishes 
as the temperature goes to zero and has a maximum when the temperature is
of the order of the Thouless temperature $T_{c}=E_{c}/k_B$. 
At elevated temperatures, $T\gg T_{c}$, the parameter controlling the 
decrease in amplitude of the conductance oscillations is $E_{c}/k_B T$. This
is simply the relative number of electrons with energy of order 
$E_c$. These electrons are responsible for the interference 
effect we are discussing, which is nothing but the ``thermal effect'' of
Refs.~\onlinecite{Nazarov Stoof} and \onlinecite{Volkov Lambert}. 

Now we turn to structures with beam splitters; below we show
that beam-splitting scattering between different trajectories at the
junctions between the mesoscopic sample and the reservoirs 
qualitatively changes the interference pattern.
In this case quasiparticles with low excitation energies, $E<E_{c}$, may
contribute --- in some cases resonantly --- to the interference effects 
causing the conductance to
oscillate as a function the superconducting phase difference.

\subsubsection{Giant conductance oscillations in Andreev interferometers
               containing beam splitters}

Scattering due to potential barriers or geometrical features 
at junctions between the mesoscopic region and the 
reservoirs qualitativly change the nature of quasiparticle trajectories. 
In particular, a particle reflected from an N-S boundary does not necessarily 
leave the sample for the reservoir directly. Instead, it may be reflected
by the junction and re-enter the mesoscopic region. There is
a certain probability that such 
reflections  creates low energy trajectories that  connect the 
reservoir(s) with both superconductors. An example of such a trajectory is 
shown in Fig.~\ref{fig22}.

The role of beam splitters in Andreev interferometers was
first payed attention to by Nakano and 
Takayanagi. \cite{Takayan} A number of other interference phenomena also
involving quasi-particles at the Fermi energy (zero temperature phenomena)
has been discussed in the literature. For instance,
Wees {\em et al.} \cite{Wees} showed that
elastic scatterers generate muliple reflections at the N-S boundary
resulting in an enhancement of the conductance above its classical value. 
In ballistic structures resonant tunneling through Andreev energy levels 
coinciding with the Fermi level was predicted in 
Refs.~\onlinecite{go,Claughton}.  
For diffusive structures containing  beam splitters a significant increase of the 
Aharonov-Bohm oscillations of the conductance  was shown to exist in 
Refs.~ \onlinecite{Naz,Zaitsev,Lambert,Claughton,Allsopp,Claughton Lambert,Volkov Zaitsev}. 
Beenakker {\em et al.} \cite{Beenakker} showed that the angular distribution 
of quasi-particles Andreev reflected by a disordered normal-metal - 
superconductor junction 
has a narrow peak centerd around the angle of incidence. The peak is higher 
than the 
coherent backscattering peak in the normal state by a large factor $G/G_0$ 
($G$ is the conductance of the junction and $G_0=2e^2/h$). The authors 
identified the enhanced backscattering as the origin  of the increase of 
the oscillation amplitude predicted in Refs.~ \onlinecite{Zaitsev,go}. As a final
example we note that it was shown in Ref.~\onlinecite{Allsopp} that the beam 
splitter violates the ``sum rule" 
according to which the conductance in the absence of junction scattering 
is equal 
to the number of transverse modes and does not depend on the superconducting 
phase difference.

All the mentioned interference phenomena involving
quasiparticles at the Fermi level ($E=0$) have the same 
nature for both ballistic and diffusive structures. 
This follows from the complete phase conjugation of electron- and hole 
excitations at the Fermi energy. At the Fermi energy even a random-walk-type 
of diffusive 
electron trajectory caused by impurity scattering is completely 
reversed by the Andreev reflected hole and there is a complete compensation
of phase. 
In particular the giant oscillations of conductance with phase difference
$\phi$ is
insensitive to impurities, as there is a finite scattering volume 
in which the phase gains along the electron-hole trajectories are 
completely compensated.  

When the transparency associated with junction scattering has intermediate 
values  both the thermal effect and the resonant oscillation effect contribute
simultaneously provided the temperature is close to the Thouless temperature. 
In experiments measuring the conductance oscillations for structures with beam 
splitters 
\cite{Petr1,Poth1,Wees2,Vegvar,Petrashov,Charlat96,Hartog96,Rahman,Ohki96,Weesbook,Takayanagi95a} 
the temperature was of the order of the Thouless temperature or higher, and 
hence both effects could contribute.   
The effects can be distinguished by lowering the temperature below the 
Thouless 
temperature, as then the amplitude 
of the conductance oscillations decrease in the case of the thermal effect 
(it goes 
to zero as the temperature goes to zero) while the resonant amplitude of 
the conductance increases and is maximal at zero temperature. 
Experimental evidence is just beginning to appear \cite{fot1,Petrashov97,Hideaki97}

While the role of the termal mechanism  has been  investigated in 
detail in Refs.~\onlinecite{Nazarov Stoof,Volkov Lambert}, for the giant 
conductance oscillations the role of intensity of  scattering for all types 
of scattering mentioned above (normal [non-Andreev] reflection at the 
N-S boundaries, 
junction- and impurity scattering)
 remains without a quantative description. The objective of this paper is 
to fill this gap. 
 
The paper is organized as follows: in Section \ref{formulation} we describe how
Andreev interferometers are modelled in this work; in Section \ref{coupling}, 
we develop a resonant perturbation theory to find the conductance
in the case of ballistic transport inside the sample and weak coupling of
the sample to the reservoirs. In comparison with Ref.~\onlinecite{go} we here allow
scattering between different conduction channels at the two junctions between
sample and leads to reservoirs.
%
% considering the $N_{\perp}\times N_{\perp}$ scattering
%matrix to be a random matrix and changing the number of transverse channels $%
%N_{\perp}$ from 10 to 40, and the superconducting phase difference $%
%\phi $ from 0 to $2\pi$;
%
In Section \ref{normalreflection} we in addition take into 
account the normal
reflection that accompanies the Andreev reflection of an electron (hole) at
a real normal conductor- superconductor interface, and get an explicit
analytical expression for the conductance of the system as a function of the
number of transverse channels.
For cases when it is inconvenient to get analytical results, such as when the
coupling is not weak, we present some results of numerical calculations in
Section \ref{numerical calculations}.
% 
%$N_{\perp}$, the parameter $\epsilon$
%characterizing the strength of the coupling, and the superconducting phase
%difference $\phi$; 
%
Then, in Section \ref{Feynman} we relax the condition of the sample being
in the ballistic  transport regime and calculate the
giant  conductance oscillations for a diffusive hybrid S-N-S structure using the 
Feynman path integral approach for  the transition probability amplitude. In
the conclusions, Section \ref{Conclusions}, we discuss the range of parameters
for which the conductance oscillations can be giant in real experiments.

	\section{Model for an Andreev interferometer}\label{formulation}
In this Section we describe our model for an Andreev interferometer. As
schematically shown in Fig.~ref system it
consists of a superconductor-normal (semiconductor)-superconductor 
(S-N-S) sample
coupled to two normal electron reservoirs between which a voltage bias is
applied. Appealing to experiments \cite{Takayanagi95a,Nitta,Nguyen,Dimoulas,Lenssen}
we neglect scattering of electrons by impurities inside the sample for the time
being and return to this point in Section \ref{Feynman}.
Nevertheless, the junctions between the S-N-S sample and the normal leads
to the electron reservoirs inevitably are sources of scattering.  
%
%Indeed, the
%junction can be considered as a part of the sample the width of which goes
%to infinity at the junction. Hence it follows that the distance between the
%electron energy levels decreases with approach to the junction that results
%in violation of the adiabatic condition and one should expect channel mixing
%and scattering of electrons at the junction.
%
So, whereas we consider electron transport to be adiabatic inside the sample
--- the current being carried in $N_\perp$ channels (modes) --- electrons can be
scattered between different conduction channels at these junctions. Taking
this into account amounts to a first generalization of our earlier treatment of
this problem. \cite{go} %
%Therefore in addition to the
%problem formulated in \cite{go} a problem of non-adiabatic scattering of
%electrons at the junctions arises. We study that effect in the next section
%using the model 
%
In our model the coupling between the sample and the reservoirs is
controlled by potential barriers (beam splitters, see above) appearing at the 
junctions between the leads from the reservoirs and the sample.
We assume that in the case of low barrier 
transparency the approximation of a nearly isolated sample
is adequate and that channel mixing is absent; we shall then study what happens
when the coupling increases in Section~\ref{coupling}.

\begin{figure}%%%[hbt]\hskip 1cm\hbox{
\centerline{\psfig{figure=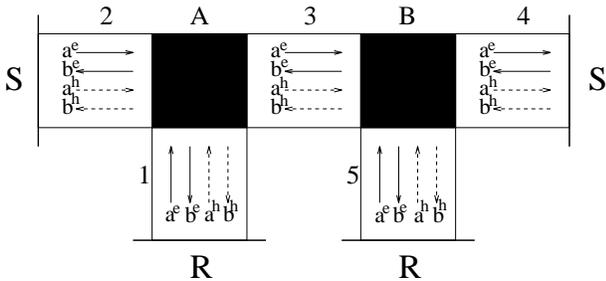,width=8cm}}
\vspace*{0.5cm}
\caption{\protect{\label{system}}
Schematic picture of an Andreev interferometer of the same
type as shown in Fig.~1. The full [dashed] arrows indicate electrons
(e) [holes (h)] moving in the ballistic segments 1-5 of the sample. 
In the model
calculation described in the text Andreev- and/or normal scattering 
may occur at the
two superconducting mirrors (S) and scattering between different segments and
channels (modes) may occur at the junctions marked A and B.
}
%%%\label{fig2a}}\vskip .5cm
\end{figure}

Another fact ignored in our earlier work \cite{go} is the possible 
``normal" reflection of quasiparticles by a Schottky barrier at the S-N boundaries, a
mechanism that would compete with the Andreev reflection. When normal reflection is possible
the degeneracy of the quasiparticle energy levels (Andreev levels) which 
occurs 
at the Fermi level for certain values of the phase difference between the two
superconductors is lifted and the ``giant" conductance oscillations as a
function of this phase difference is greatly reduced.
Despite the experimental fact that the 
probability for normal reflection is small \cite{Takayanagi95a,Nitta,Nguyen,Dimoulas,Lenssen} 
the criteria for
how small the normal reflection probability must be for the giant oscillations
to survive is obviously  an important question, which we consider in Section
\ref{normalreflection}. Below we formulate our transport problem for a general
case which includes both the possibility for scattering between conduction
channels at the sample-lead junctions and normal reflections at the N-S
interfaces.  
%
%The analysis of the
%general case is presented in Section~\ref{normalreflection} where the normal
%part is 
%considered as diffusive. Using the Feynmann path integral approach we study the 
%probabilities of reflection and transmission of quasiparticles via
%the sample and show the %giant oscillation effect to survive.

\begin{figure}
\centerline{\psfig{figure=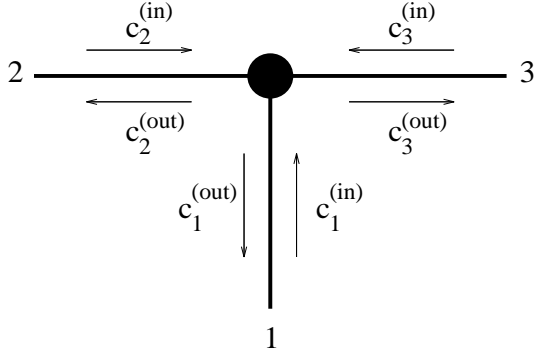,width={7 cm}}}
\vspace*{0.5cm}
\caption{\protect{\label{junction}}
Detail of the junction (A, B in Fig.~4) coupling the 
reservoirs via leads to the normal part of the 
system. A scattering matrix connects the amplitudes of incoming- and outgoing
quasiparticles, see text.
}
\end{figure}

It is convenient to divide our model Andreev interferometer into five different segments so
that the electron transport to a good approximation is adiabatic in each segment. We then
use a phenomenological method for describing the two manifestly non-adiabatic junction
regions [marked A and B in Fig.~ref system]. The
quasi-particle wave functions in the adiabatic segments 1-5 shown in Fig.~\ref{system} can
be found with the help of the Bogoliubov-de-Gennes equation. As channel mixing is absent in
the adiabatic segments the electron- and hole like components of the wave function in the
$n$:th transverse mode in segment $\alpha$ is   
\begin{eqnarray}
\nonumber
u_\alpha (x,y)=\sum_{n=1}^{N_{\perp}}\left( a_{\alpha
,n}^{(e)}e^{ik_n^{(e)}x}+b_{\alpha ,n}^{(e)}e^{-ik_n^{(e)}x}\right)\sin k_{\perp }(n)y \\
v_\alpha (x,y)=\sum_{n=1}^{N_{\perp}}\left( a_{\alpha
,n}^{(h)}e_n^{-ik^{(h)}x}+b_{\alpha ,n}^{(h)}e_n^{ik^{(h)}x}\right)\sin k_{\perp }(n)y
\label{wf}
\end{eqnarray}
Here $a_{\alpha ,n}^{(e)}$ and $b_{\alpha ,n}^{(e)}$ [$a_{\alpha ,n}^{(h)}$
and $b_{\alpha ,n}^{(h)}$] are the probability amplitudes  for free motion of
electrons [holes] forward and backward, 
respectively, in channel $n$ and segment
$\alpha $ of the sample; $k_{\perp }(n)=n\pi/d, n=0, 1, 2,\ldots$ is
the quantized transverse wavevector assuming a hard wall confining potential, $d$ is the
width of the sample, $k^{(e,h)}_n=[k_F^2-k_{\perp}^{2}(n)\pm 2mE/\hbar^2]^{1/2}$ is the
electron (hole) longitudinal momentum, $k_F$ is the Fermi wavevector, $E$ is the electron
energy measured from the Fermi energy, while $x$ and $y$ are longitudinal and transverse
coordinates in the sample, respectively. Non-adiabatic scattering of electrons in the
junction regions, see Fig.~\ref{system}, is described by a unitary scattering matrix
$\hat S$  connecting the wave functions in the surrounding sample segments. Scattering at
these junctions mix the transverse modes (channels) in the adiabatic segments (which here
and below, for the sake of simplicity, are considered to have the same number of open
transverse channels).  Hence, the scattering
matrix connects ${\bf c}_\alpha ^{(in)}$ and ${\bf c}_\alpha ^{(out)}$, which are $N_{\perp}$
-component vectors whose coefficients $a_{\alpha,n}^{(e,h)}$, $b_{\alpha ,n}^{(e,h)}$
describe the incoming and outgoing adiabatic wave functions [see Fig.~\ref{junction} and
Eq.~(\ref{wf})], % 
\begin{equation}
{\bf c}_\alpha ^{(out)}=\sum_{\beta =1}^3{\bf \hat S}_{\alpha \beta }{\bf c}
_\beta ^{(in)}  
\label{GenS}
\end{equation}

We assume the coupling matrix ${\bf \hat S}$ to be symmetric with respect to
the left and right sample segments [labeled 2 and 4 
in Fig.~\ref{system}]. Therefore the matrix can be written as
\begin{equation}
{\bf \hat S}=\left(
\begin{array}{ccc}
{\bf \hat s}_{11} & {\bf \hat s}_{12} & {\bf \hat s}_{12} \\
{\bf \hat s}_{12} & {\bf \hat s}_{22} & {\bf \hat s}_{23} \\
{\bf \hat s}_{12} & {\bf \hat s}_{23} & {\bf \hat s}_{22} \\
\end{array}
\right) \; ,
\label{Gens}
\end{equation}
where ${\bf \hat s}_{\alpha \beta }$ are $N_{\perp}\times N_{\perp}$
matrices which mix the conduction channels when an electron (or hole) 
is transferred from the $\beta$ - to the $\alpha$ segment. 
Electrons and holes are, however, not mixed. The
elements of ${\bf \hat s}_{\alpha \beta }^{nm}$ ($n,m=1,2,...N_{\perp}$) are the
probability amplitudes for an electron (or hole) in the $m$-th channel of the
$\beta$-section to be transferred to the $n$-th channel of the $\alpha$-section. We assume
that scattering of an incident quasiparticle at the junction causes transmission of the
electron into each of the $N_{\perp}$ open transverse channels with a probability, which is
of the same order of magnitude for all channels. This implies that the matrix elements of 
the matrices $\hat s_{12}$, $\hat s_{22}$, $\hat s_{23}-1$ and $\hat s_{11}-1$ 
 are of order $1/\sqrt{N_{\perp}}$. 

\noindent
We choose to parametrize the
${\bf \hat S}$ matrix in a way such that there is no channel mixing if the sample is
completely decoupled from the reservoirs. This coupling is determined by the
elements of the matrix ${\bf \hat s}_{12}$ which are the probability amplitudes for 
electron (hole) transitions from a lead (segments 1 or 5) to the sample (segments 2, 3 or
3,4).
In order to describe the strength of the coupling we introduce the parameter
$\epsilon_r$ and write
\begin{equation} 
{\bf \hat s}_{12}=\left(\frac{\epsilon_r}{N_{\perp}}\right)^{1/2}{\bf \bar s}_{12} \; .
\label{eps} 
\end{equation}
 The scattering matrix
(\ref{Gens}) has to be unitary --- a requirement that leads to five relations
between the eight matrices
${\bf \hat s}_{\alpha \beta }$, see Eqs.~(\ref{uncon1}-\ref{uncon5}) 
in Appendix~A. (Note that each matrix has an independent Hermitian and
anti-Hermitian part). We are thus left with three undetermined matrices, which we choose to
be ${\bf \hat s}_{12}$ and the anti-Hermitian part of ${\bf \hat s}_{22}$ (the last choice
is made only for the sake of calculational convenience, see Appendix~A.
 We assume the Hermitian and anti-Hermitian parts of ${\bf \hat
s}_{22}$ to be of the same order in the parameter $\epsilon_r$ since in the general
case they are connected by a Kramers-Kronig relation. It follows
from the unitarity conditions that the matrix elements of ${\bf \hat s}_{12}$ are of
order $1/\sqrt{N_{\perp}}$. Hence the matrix elements of ${\bf \bar s} _{12}$ are
of order unity.

 The conductance of our model system is in the limit of vanishing bias voltage
determined by the Landauer formula as modified by Lambert for
a system with Andreev reflections \cite{Lambert1}:
\begin{equation}
G=\frac{2e^2}h\left( T_0+T_A+\frac{2(R_AR_A^{\prime }-T_AT_A^{\prime })}{%
T_A+T_A^{\prime }+R_A+R_A^{\prime }}\right) \; .
\label{gencond}
\end{equation}
Here $e$ is the electron charge, $h$ is Planck's constant and
\begin{eqnarray} \nonumber
T_A=\sum_{k=1}^{N_{\perp}}\tau _k^{(A)}, & & R_A=\sum_{k=1}^{N_{\perp}}\rho
_k^{(A)}\\
T_0=\sum_{k=1}^{N_{\perp}}\tau _k^{(0)}, & & R_0=\sum_{k=1}^{N_{\perp}}\rho
_k^{(0)}. 
\label{probab}
\end{eqnarray}
In Eq.~(\ref{probab}) $\tau _k^{(0)}$ [$\tau _k^{(A)}$] 
is the probability for an electron approaching the sample in the $k$-th
transverse channel of the left lead to be transmitted as an electron (hole)
into any of the outgoing channels of the right lead.
The quantity $\rho _k^{(0)}$ 
[$\rho _k^{(A)}$] is the probability for the same electron to be reflected as an
electron (hole) into any outgoing channel in the same left lead it came from.
Similarily, $\tau _k^{{\prime}(0)}$ [$\tau _k^{{\prime} (0)}$] and $\rho_k ^{{\prime
}(0)}$ [$\rho_k ^{{\prime} (A)}$] are normal (Andreev)
probabilities for an incoming electron from the $k$-th transverse channel of
the right lead to be transmitted as an electron (hole) into any outgoing
channel of the left lead and to be reflected as an electron (hole) back into any
outgoing channel of the right lead, respectively.

In order to proceed we have to solve the matching
equations for the adiabatic wave functions in sample and leads.
The matching problem under consideration is illustrated in Fig.~\ref{system} 
where solid and dashed arrows symbolically show electron- and hole plane waves
moving to the right and to the left, respectively. The coefficients ${\bf
a}_\alpha ^{(e,h)}$ and ${\bf b}_\alpha ^{(e,h)}$ are $N_{\perp}$-component vectors, the
components of which are the probability amplitudes $a_{\alpha ,n}^{(e,h)}$ and $b_{\alpha
,n}^{(e,h)}$, see Eq.~(\ref{wf}). Matching the wave functions at the junctions
using Eq.~(\ref{GenS}) one gets the following set of equations for these
amplitudes: 
\begin{equation} 
\left\{
\begin{array}{l}
{\bf b}_1^{(e,h)}={\bf \hat s}_{11}{\bf a}_1^{(e,h)}+{\bf \hat s}_{12}{\bf a}
_2^{(e,h)}+{\bf \hat s}_{12}{\bf b}_3^{(e,h)}; \\
{\bf b}_5^{(e,h)}={\bf \hat s}_{11}{\bf a}_5^{(e,h)}+{\bf \hat s}_{12}\hat
U_3^{(e,h)}{\bf a}_3^{(e,h)}+{\bf \hat s}_{12}\hat U_4^{(e,h)}{\bf b}
_4^{(e,h)} \\
\end{array}
\right.   
\label{match1}
\end{equation}
and
\begin{equation}
\left\{
\begin{array}{l}
{\bf b}_2^{(e,h)}-{\bf \hat s}_{22}{\bf a}_2^{(e,h)}-{\bf \hat s}_{23}{\bf b}
_3^{(e,h)}={\bf \hat s}_{12}{\bf a}_1^{(e,h)}; \\
{\bf a}_3^{(e,h)}-{\bf \hat s}_{23}{\bf a}_2^{(e,h)}-{\bf \hat s}_{22}{\bf b}
_3^{(e,h)}={\bf \hat s}_{12}{\bf a}_1^{(e,h)}; \\
\hat U_3^{(e,h)\dag }{\bf b}_3^{(e,h)}-{\bf \hat s}_{22}\hat U_3^{(e,h)}{\bf a}
_3^{(e,h)}-{\bf \hat s}_{23}\hat U_4^{(e,h)\dag }{\bf b}_4^{(e,h)}={\bf \hat s}
_{12}{\bf a}_5^{(e,h)}; \\
\hat U_4^{(e,h)\dag}{\bf a}_4^{(e,h)}-{\bf \hat s}_{23}\hat U_3^{(e,h)}{\bf a}
_3^{(e,h)}-{\bf \hat s}_{22}\hat U_4^{(e,h)}{\bf b}_4^{(e,h)}={\bf \hat s}
_{12}{\bf a}_5^{(e,h)};
\end{array}
\right.   
\label{match2}
\end{equation}
Here the diagonal matrices $\hat{U_\alpha}^{(e,h)}$ simply keep track of the
phase gained by electrons and holes during their free motion across segment
$\alpha$. The diagonal matrix elements are
\begin{equation}
\begin{array}{ccc}
u_\alpha^{(e)}(n)=e^{ik^{(e)}_n l_\alpha}, & 
u^{(h)}_{m}(n)=e^{-ik^{(h)}_n l_\alpha}, & n=1, 2,\ldots N_\perp
\end{array}
\label{phase}
\end{equation}
where $l_\alpha$ is the length of section $\alpha$ in Fig.~\ref{system}.

The set of equations (\ref{match1}) and (\ref{match2}) must be
supplemented with boundary conditions at the N-S interfaces. In the general
case when both Andreev- and normal reflections at the N-S boundaries are
possible the boundary conditions are 
\begin{eqnarray}
\nonumber
a_{2,n}^{(e)}&=&e^{i\Psi_{1}}
\left[r_N^{(1)}e^{i2k^{(e)}(n)l_2}b_{2,n}^{(e)}+r_A^{(1)}e^{i\delta
k(n)l_2}b_{2,n}^{(h)}\right]   \\
a_{2,n}^{(h)}&=&e^{i\Psi_1}\left[-r_A^{(1)*}e^{i\delta
k(n)l_2}b_{2,n}^{(e)}+r_N^{(1)*}e^{-i2k^{(h)}(n)l_2}b_{2,n}^{(h)}\right]
\label{andnorm1}
\end{eqnarray}
for the left (first) boundary  [$\delta k\equiv k^{(e)}-k^{(h)}$] and
\begin{eqnarray}
\nonumber
b_{4,n}^{(e)}&=&e^{i\Psi
_2}\left[r_N^{(2)}a_{4,n}^{(e)}+r_A^{(2)}a_{4,n}^{(h)}\right]  \\
b_{4,n}^{(h)}&=&e^{i\Psi_2}\left[-r_A^{(2)*}a_{4,n}^{(e)}+
r_N^{(2)*}a_{4,n}^{(h)}\right] 
\label{andnorm2}
\end{eqnarray}
for the right (second) boundary, see Fig.~\ref{system}. The probability amplitudes
for normal- and Andreev reflections at the N-S boundary are given by $e^{i\Psi
}r_N$ and $e^{i\Psi }r_A $. It follows that
$|r^{(1,2)}_N|^2+|r^{(1,2)}_A|^2=1$. \cite{Blonder82} For
convenience explicit expressions for these quantities in terms of 
the complex order parameter of the superconductor and the 
reflection- and transmission probability amplitudes 
of the normal barrier at the N-S interface are given in Appendix~B.

The equations (\ref{match1}), (\ref{match2}), (\ref{andnorm1}), and (\ref{andnorm2}) 
together with the conductance formula 
(\ref{gencond}) form a complete set of equations that permits us to find the
conductance of the system under consideration.

In the next section we discuss the non-adiabatic scattering of electrons at
the junctions and present analytical formulae for the case of a weak
coupling of the sample to the reservoirs ($\epsilon_r\ll 1$) and numerical
results of computer simulations in the general case. The role of the
non-Andreev (normal) reflection at the N-S boundaries is discussed in
section \ref{normalreflection}.
	\section{Role of scattering and mode mixing at the points of
 coupling to the reservoirs}\label{coupling}
We start our analysis by assuming a weak coupling between sample and
(leads to) reservoirs. 
In this case the parameter $\epsilon_r$ introduced in Eq.~(\ref{eps}) is
much smaller than one. It is convenient to develop a qualitative understanding starting from
the so called Andreev levels that form in the isolated sample when $\epsilon_r$ is strictly zero.
We consider values of $\phi$ near odd multiples of $\pi$ for which Andreev levels will appear
at the Fermi energy ($\phi =\phi_2-\phi_1$; $\phi_1$ and $\phi_2$ are the phases of the gap
functions in the left and right superconductor, see Fig.~\ref{system}).  We concentrate on energies
in a narrow interval $\Delta E\sim \epsilon_r\hbar v_F/L$ around the Fermi energy,  whithin which
the quantum states of electrons perturbed by a coupling of the sample to the
reservoirs are expected to be found  ($\hbar v_F/L$ is the
characteristic spacing of mode energy levels near the Fermi energy).

%There will be $N_R\leq N_{\perp}$ energy levels of the transverse quantization of the
%electron motion of the isolated sample. In the situation considered in this section (electrons and
%holes undergo only Andreev reflections at the N-S boundaries) $N_R=N_{\perp}$ as was discussed
%above. But in the case of Andreev + normal reflections at the N-S boundaries considered in the next
%section the degeneration is taken away and $1\ll N_R\leq N_{\perp}$.

By solving the matching equations (\ref{match2}), (\ref{andnorm1}), and (\ref{andnorm2}) for 
$\epsilon_r=0$, using Eqs.~(\ref{nandr1}),
(\ref{nandr3}), and (\ref{nandr4}) of Appendix~B one recovers the following
expression \cite{Kulik}  for the low
energy spectrum of electron-hole quasiparticles in the normally conducting semiconductor 
sandwiched between two superconducting mirrors
\begin{eqnarray}
\label{Aspectr}
&&
\pi (2l+1)\pm \phi = \left(\sqrt{k_F^2-k_{\perp}(n)^{2}+2mE/\hbar^2}-\right.\\
&& \quad\quad\left.\sqrt{k_F^2-k_{\perp }(n)^{2}-2mE/\hbar^2}
\right)L, \quad 
l=0,\pm 1,\pm 2,\ldots
\nonumber
\end{eqnarray}
Here $m$ is the electron mass, $L=l_2+l_3+l_4$ is the length of the normal part of the sample. 
and all reflections are assumed to be of the Andreev type.
The phase $\phi$ comes with a plus- or a minus sign in Eq.~(\ref{Aspectr})
depending on whether the electron-hole excitations move as electrons or holes when going
from the left to the right S-N interface.

Expression (\ref{Aspectr}) for the spectrum  tells us that when $\phi =\pi (2l^\prime+1)$ 
there is one Andreev state at the Fermi level for each  transverse mode (index $n$)
simultaneously, i.e., the energy of the
state whose quantum number $l$ associated with the longitudinal motion equals $l^\prime$ coincides
with the Fermi energy irrespective of mode number $n$.
Therefore, the degeneracy of the energy level at the Fermi energy ($ E=0$) is given by the number of
open transverse modes $N_{\perp}$, whenever $\phi$ equals an odd multiple of $\pi$. This results in
a giant probability for resonant transmission of electrons from one reservoir to the other. The
amplitude of the corresponding conductance oscillations, $\Delta G\propto N_{\perp} e^2/\hbar $
\cite {go}, is therefore much larger than the conductance quantum. 

A finite coupling of the sample to the
reservoirs (which is of course necessary for a current to be observable) simultaneously results in
a broadening and a shift of the Andreev energy levels. The former effect is due to quasiparticle
tunneling from the sample to the reservoirs after a finite time, the latter is due to mixing of the
transverse modes that inevitably accompanies a finite coupling. 
Below we show the broadening and the
shift to be of the same order in the transparency of the barrier connecting sample and
reservoirs. The result is a broadening of the peaks of resonant sample conductance
but their giant amplitude remains. This is beacuse a Breit-Wigner type of resonance is
broadened without loss of amplitude when the coupling is increased. It turns out that the
broadening of each state tends to compensate the shifting around of the energies of previously
degenerate states. Readers who are not interested in technical details may want to turn directly to
Eq.~(\ref{G}), which expresses this result.  Results of numerical calculations presented in Section~\ref{numerical calculations} show this picture to hold up to a value for $\epsilon_r$ which is about half its
maximum value. A further increase of the coupling results in a large decrease of the amplitude and
increase of the broadening of the peaks.

In the weak coupling case the set of equations (\ref{match1}), (\ref{match2}), (\ref{andnorm1}),
and (\ref{andnorm2}) which determines the transmission probability amplitudes can be
solved analytically
 by perturbation theory in the small parameter $\epsilon_r$. Below this perturbation theory will be
developed.

As shown in Appendix~A, all the matrices which
describe scattering of electrons and holes inside the sample ($\hat s_{22}$
and $\hat s_{23}-\hat 1$) are proportional to $\epsilon_r$ if the coupling
matrix $\hat s_{12}$ is proportional to $\sqrt{\epsilon_r}$ and if $\epsilon_r\ll 1$ . Hence it
follows that the set of equations (\ref{match1}-\ref{andnorm2}) can be written in the form
\begin{equation}
\left( {\bf \hat W}(E)-\epsilon_r{\bf \hat \Omega }\right) |{\bf H}\rangle=
{\left(\frac{\epsilon_r}{N_{\perp}}\right)}^{1/2}|{\bf K}\rangle \; . \label{main}
\end{equation}
The vector $|{\bf H}\rangle$ has $12N_{\perp}$ unknown components, viz. 
${\bf a}_i^{(e,h)},$ ${\bf b}_i^{(e,h)}$ for $i=$2, 3, and 4. The vector $|{\bf K}\rangle$ has
$4N_{\perp}$ known elements, ${\bf a}_k^{(e,h)}$ for $k=$1 and 5, and $8N_{\perp}$
elements which are zero. The matrix ${\bf \hat W}(E)$ has $12N_{\perp}$ block matrices
along the diagonal with non-zero elements % 
\begin{equation} 
[{\bf W}_{\alpha ,\beta }(E)]_{nm}=
\delta_{nm}{\bf w}_{\alpha \beta }^{(n)}(E) \quad ,
\label{W} 
\end{equation}
where $\alpha ,\beta =1,2,\ldots 12$; $n,m=1\ldots N_{\perp}$. The 
matrix ${\bf \hat w}^{(n)}(E)$ has been obtained for the $n$-th fixed 
transverse 
mode in the absence of coupling ($\epsilon_r=0$) by matching the electron- and
hole components of the wave functions at the N-S boundaries using Eqs.~(\ref{andnorm1}) and
(\ref{andnorm2}) and at the junctions coupling the sample to the electron reservoirs
using Eq.~(\ref{match2}) for fixed channel number $n$ and $\epsilon_r=0$. The matrix
${\bf\hat\Omega}$ has elements describing mixing between modes. The explicit forms of
operators $\hat{\bf W}, \hat{\bf\Omega}$ and vectors $|{\bf H}\rangle, |{\bf K}\rangle$ are
straightforwardly found by comparing Eqs.~(\ref{match2}) and (\ref{main}).

In order to use the resonant perturbation theory we have to consider some
properties of the unperturbed system relevant to our problem. It is
straightforward to see from (\ref{W}) that the determinant of the matrix ${\bf \hat W}(E)$
can be written as a product of $N_\perp$ factors
\begin{equation}
\mbox{Det}{\bf \hat W}(E)=\mbox{Det}{\bf \hat w}^{(1)}(E)
\times\ldots\times \mbox{Det}{\bf \hat w}^{(N_{\perp})}(E)  \label{detprod}
\end{equation}
and that its value iz zero at any eigenvalue $E=E_{n,l}$ of the unperturbed system. The 
eigenfunctions $|\psi_l ^{(n)}\rangle$ of the unperturbed problem satisfy the following equation
\begin{equation}
{\bf \hat W}(E_{n,l})|\psi_l^{(n)}\rangle=0
\label{feq2}
\end{equation}
Developing the
perturbation theory we assume the following inequality to be satisfied.
\begin{equation}
\lambda _F\ll d\ll L \; ,
\end{equation}
where $\lambda _F$ is the de-Broglie wave length (Fermi wave length) 
of the electron, while $d$
and $L$ is the width and length of the sample. We note that the 
perturbation of the
energy has to be much smaller
 than the distance between neighbouring energy levels
corresponding to quantization of the longitudinal motion of electrons, that is
\begin{equation}
\epsilon_r\hbar v_F/L\ll \hbar v_F/L \; . \label{enineq}
\end{equation}
Here we develop the perturbation theory for a general case in order
to use the results also in the next section. Therefore in order to find the correct
zero order wave-function the vector $|{\bf H}\rangle$ must be taken
as a superposition of the $N_R$ states inside the resonant region ($N_R$ is possibly but not
necessarily smaller than $N_\perp$) 
\begin{equation}
|{\bf H}\rangle=\sum_{n=1}^{N_R}\gamma _n|\psi_l^{(n)}\rangle+|{\bf H}_1\rangle
\; .
\label{pwf}
\end{equation}
The summation in (\ref{pwf}) goes over the $N_R$ transverse modes inside the resonant region, 
which extends over an interval of order
 $\epsilon_r\hbar v_F/L$ on either side of the Fermi energy; $|{\bf H}_1\rangle
$ is a small addition $\propto \epsilon_r$. The unknown coefficients 
$\gamma _n$ should be found with the solvability condition of the equation
for $|{\bf H}_1\rangle$ that is readily available from Eq.~(\ref{main}) in the
linear approximation in $\epsilon_r\ll 1$:
\begin{eqnarray}
\nonumber
{\bf \hat W}(E)|{\bf H_1}\rangle=&-&\sum_{n=1}^{N_R}
\gamma _n[{\bf \hat W}^{\prime
}(E_{n,l})(E-E_{n,l})-\epsilon _r{\bf \hat \Omega }]|
\psi_l^{(n)}\rangle\\
&+&{\left(\frac{\epsilon_r}{N_{\perp}}\right)}^{1/2}|{\bf K}\rangle
\label{faf}
\end{eqnarray}
Here the superscript ``prime'' indicates derivation with respect to energy $E$. 
When obtaining
Eq.~(\ref{faf}) we used the inequality (\ref{enineq}) and expanded ${\bf \hat W}(E)$ in a
Taylor series  around $E_{n,l}$ (with the restriction $|E-E_{n,l}|\ll \hbar v_F/L$) 
in every term of the sum and took into account Eq.~(\ref{feq2}).

Multiplying both sides of Eq.~(\ref{faf}) from the left by bra-vectors 
$\langle \psi_l^{(m)}|$ [which can be determined from the equation $\langle\psi_l^{(m)}|{\bf \hat
W}(E_m)=0$] one readily gets the solvability conditions for Eq.~(\ref{faf}) that determines
the coefficients $ \gamma _n$. In this way we obtain the main equation which has
to be solved in order to
 get $\gamma_n$; these coefficients, according to Eqs.~(\ref{match1}) and (\ref{pwf}),
determine the probability of the resonant transmission of an electron from
one reservoir to the other via the sample):
\begin{equation}
\sum_{n=1}^{N_R}\left[ i(E-E_n)W_n^{\prime }\delta _{mn}-\epsilon_r\Omega
_{mn}\right] \gamma _n=\left( \frac{\epsilon_r}{N_{\perp}}\right)
^{1/2}K_m \; .
\label{major}
\end{equation}
Here we have used the short hand notation
\begin{equation}
W_n^{\prime }=-i\langle{\bf \psi }^{(n)}|{\bf \hat W}^{\prime }(E_n)|{\bf \psi }
^{(n)}\rangle  \label{Wderiv}
\end{equation}
\begin{equation}
\Omega_{mn}=\langle{\bf \psi }^{(m)}|{\bf \hat \Omega }|{\bf \psi }^{(n)}\rangle
\label{Omel}
\end{equation}
\begin{equation}
K_m=\langle{\bf \psi }^{(m)}|{\bf K}\rangle
\label{Kel}
\end{equation}
We have also dropped the subscript $l$ as we have assumed it
does not change under the perturbation considered. Using Eq.~(\ref{feq2})
for $\epsilon_r=0$ it is straightforward to calculate $W_n^{\prime }$ and
show it to be real, i.e., the Hermitian and anti-Hermitian parts of the
coupling matrix ${\bf \hat \Omega }$ provide broadening and shift
of the energy levels of the sample, respectively. 
In our analysis of Eq.~(\ref{major}) we consider the matrix elements $\Omega
_{nm}$ to be of order unity. \cite{note_order}
It is then easy to see that far from the resonance, where $\hbar v_F/L\gg|E-E_n|\gg
\epsilon_r\hbar v_F/L$, the first term on the right hand side of Eq. (\ref{major}) dominates and
one gets 
\begin{equation}
 \gamma_n\approx \left( \frac{\epsilon_r}{N_{\perp}}\right)
^{1/2}\frac{K_n }{iW_n^{\prime }(E-E_n)}\quad  .
\label{outres} 
\end{equation} 
Knowing $\gamma_n$ we may
calculate $|{\bf H}\rangle$, which contains
the coefficients $a_3$ and
$b_4$, from Eq.~(\ref{pwf}). By using
Eq.~(\ref{match1}) the probability of transmission of an electron (hole) from one reservoir to the
other is  \begin{equation} 
|b_5^{(e,h)}|^2\sim \epsilon_r^2\quad  
\label{outres2}
\end{equation} 
In the range of resonant energies, $|E-E_n|\leq \epsilon_r\hbar v_F/L$, the
amplitudes $\gamma _n\sim 1/\sqrt{N_{\perp}\epsilon_r}$ are much larger and, therefore, the
transmission probability amplitudes
\begin{equation}
b_{5,m}^{(e,h)}=\sum\limits_{n=1}^{N_R}s_{12}^{mn}
\gamma _n(e^{\pm ik_n^{(e,h)}l_3}a_{3,n}^{(0)(e,h)}+e^{\pm ik_n^{(e,h)}l_4}b_{4,n}^{(0)(e,h)})
\label{gamosc}
\end{equation}
obtained from Eqs.~(\ref{match1}), (\ref{pwf}), and (\ref{eps}) are independent
of $\epsilon_r$. Note that $a_{3,n}^{(0)(e,h)}$ and $b_{4,n}^{(0)(e,h)}$
are the known amplitudes of the wave function of the electron (hole) in
the $n$-th transverse mode in sample segments 3 and 4 when isolated
from the reservoirs. Hence $a_{3,n}^{(0)(e,h)}$ and $b_{4,n}^{(0)(e,h)}$ are of
order unity and it follows that the probability for an electron in the $m$-th
transverse mode of segment 1 --- the lead from the left reservoir --- to be transmitted to any of
the $N_{\perp}$ transverse modes in segment 5 --- the lead to the right reservoir, (see Fig.~\ref{system}) --- via the sample is 
\begin{equation}
T_{(e,h)}^{(m_0)}=\sum\limits_{m=1}^{N_{\perp }}|b_{5,m}^{(e,h)}|^2\sim
\sum_{m=1}^{N_{\perp}}\sum_{n=1}^{N_R}|s_{12}^{mn}|^2|\gamma
_n^{(m_0)}|^2\sim \frac{N_R}{N_{\perp}}\quad .
\label{trnp}
\end{equation}
The last similarity relation follows since $\sum_{k=1}^{N_{\perp}}|s_{12}^{mn}|^2\sim
\epsilon_r$ and since in the resonance region, according to Eq.~(\ref{major}), $|\gamma
_n^{(m_0)}|\sim 1/\sqrt{ N_{\perp}\epsilon_r}$ (to see this note that Eq.~(\ref{outres}) is valid
up to the resonant region where $|E-E_n| \sim \epsilon_r\hbar v_F/L$; the superscript $m_0$
indicates that the incoming electron in segment 1 moves in mode number $m_0$). Therefore in
accordance with the Landauer-Lambert formula (\ref{gencond}), the order-of-magnitude conductance in
the resonant region of a system with $N_{\perp}$ incoming electrons is
\begin{equation}
G\sim \frac{e^2}{h}\sum_{m_0=1}^{N_{\perp}}T_{(e,h)}^{(m_0)}\sim \frac{e^2}{h}N_R
\; ,  \label{G}
\end{equation}
while off the resonance $G\sim \epsilon_r^2N_Re^2/h$ [cf. Eq.~(\ref
{outres2})]. 

Since at zero temperature the energy of the incoming electrons coincides with the Fermi energy,
resonant transmission occurs in the vicinity of $\phi =\pi (2l+1),\ l=0,\pm 1,\pm 2,...$, the width
of the resonance being of order $\epsilon_r\ll 1$.
If reflections from the N-S boundaries are only of the Andreev type it follows that
$N_R$ in (\ref{G}) is equal to $R_\perp$. In this case
the conductance oscillates with $\phi $, the amplitude of
the oscillations being proportional to the total number of the transverse
modes $N_{\perp}$. In the above analysis, for the sake of simplicity, we
assumed the number of transverse modes inside the sample and the leads to be
equal but it can easily be shown that if these numbers are different the
conductance is proportional to the smallest one.

As demonstrated in this Section, for the many-channel case with mixing of transverse modes at the
junctions our analytical  approach permits us to estimate the conductance in the region far
from the resonance. It is also possible to find the width of the resonant peak and its height
(i.e., the amplitude of the conductance oscillations) but it does not permit us to find the fine
structure of the resonant peak as it is determined by the set of $N_{\perp}\gg 1 $ algebraic
equations of Eq.~(\ref{major}). Here we consider instead the fine structue of the resonant peak
using the most simple model of a one-channel sample weakly coupled to the reservoirs. In this case
calculations of the conductance in the vicinity of the resonance ($\delta\phi\equiv|\phi-\pi|\ll
1$) give the result
\begin{equation}
G=\frac{2e^2}{h}\frac{(4\gamma)^2}{[(2\gamma)^2+(\delta\phi)^2]^2}[(2\gamma \cos kl_3
+\delta\phi\sin kl_3)^2+(\delta\phi)^2] 
\label{onemode}
\end{equation} 
($k$ is the electron wave number, $l_3$ is the
distance between the junctions, $\gamma =|s_{12}|^2\sim \epsilon_r\ll 1$). It follows
that there is a dip in the middle of  the resonant peak (which appears due to an interference
between  the wave functions of the clockwise and counter clockwise motions of the quasiparticles).
When $\delta\phi=0$  the conductance is
\begin{equation} 
G=\frac{2e^2}{h}\cos^2 kl_3 \; ,
\label{pi1}
\end{equation}
and hence is it goes to zero for certain values of the wave number $k$; the
resonant peak is split into two peaks. 

In the many channel case every mode has its own longitudinal momentum, and the conductance being a 
sum over the channels is self-averaged with respect to momentum. Such an averaging of the
conductance in Eq.~(\ref{onemode}) followed by a multiplication by the number of transverse modes
gives as a result for the conductance, 
\begin{equation}
G=N_{\perp}\frac{2e^2}{h}2\gamma^2\frac{(2\gamma)^2+3(\delta\phi)^2}{[(2\gamma)^2+(\delta\phi)^2]^2}
\label{nmode}
\end{equation}
This result tells us that  there is a dip in the middle of the resonant peak with a depth of 
1/9 of the height of the resonant peak. Equation (\ref{nmode}) is valid in the absence of
transverse mode mixing. Numerical calculations of the conductance for the general case of the
transverse mode mixing also show such a dip in the middle of the resonant peak (see below).

\section{Influence of normal quasiparticle reflection at the N-S
boundaries on the giant conductance oscillations.}\label{normalreflection}
In experiments a typical N-S boundary is an interface of two different
conductors, resulting in two-channel reflection of electrons at the N-S
boundary that is an incident electron is reflected back remaining in the
state of an electron-like excitation with probability $|{ r}_N |^2$
(the normal channel) and in a state of a hole-like excitation with
probability $|{ r}_A|^2=1-|{ r}_N{ |^2}$ (the Andreev channel).  In the general case 
of nonequivalent normal barriers at the NS boundaries the quantized energy levels of an S-N-S system
are repelled from the Fermi level and the degeneracy is lifted. However, we know from
experiments  \cite{Takayanagi95a} that a situation
with a low probability for non-Andreev (normal) reflection can be realized in practice. 
Therefore it is important to derive a criterion for how low
this probablity for normal reflection must be to preserve the  giant conductance
oscillations. In this Section we discuss the role of the normal
reflections for the oscillations of the conductance in a
ballistic S-N-S system with combined Andreev and normal reflections at the
S-N boundaries.

\subsection{\label{symmetric reflection}Normal reflection from two identical barriers
at the N-S interfaces}

We start from the case of a sample isolated from the reservoirs, the
geometry of which is presented in Fig.~\ref{system},
 and assume the reflection
properties at the two NS boundaries to be identical. Matching the wave functions of the electron-
and hole-like excitations at the N-S boundaries using Eqs.~(\ref{andnorm1})
 and (\ref{andnorm2}) 
gives as a result the following spectral function, 
\begin{equation}
Q_n=\cos \varphi _{-}-|{ r_N}|^2\cos \varphi _{+}+|{ r_A}|^2\cos \phi
\quad 
\label{snsf}
\end{equation}
Here $\varphi_{-}=2mE L/\hbar^2 k_n$, $\varphi_{+}=2k_n L$, the
parallel component of the wavevector is $k_n=[k_F^2-k_{\perp}(n)^2]^(1/2)$ where $
k_{\perp}(n)$ is the projection of the wavevector on the N-S boundaries, $n$ 
labels transverse modes and $\phi$ is the phase
difference between the order parameters in the two superconductors.

For energies $E$ small compared to the energy gaps in the superconductors the equation 
%%%\begin{equation}
%
$Q_n=0$
%%%\quad 
%%%\label{dispeq}
%%%\end{equation}
%
determines the discrete Andreev energy levels of the system. This relation can be rewritten as
\begin{eqnarray}
\label{level}
E_{n,l}&=&\left[ \pi (2l+1)\pm \right. \\ 
&&\left.
\arccos\left(|{ r_A}|^2\cos \phi -|{ r_N}%
|^2\cos \varphi _{+}\right)\right] \frac{\hbar^2 k_n}{2mL}\quad ,
\nonumber
\end{eqnarray}
where the longitudinal and transverse quantum number is $l=0,\pm 1,\pm2,...$ and 
$n=1,2,...,N_{\perp}$, respectively.

In the absence of normal reflection at the N-S boundaries ($r_N=0$) Eq.~(\ref{level}) 
reduces to 
Eq.~(\ref{Aspectr}) 
and the
energy level at the Fermi energy is $N_{\perp}$-fold degenerate for values of $\phi$ that
corresponds to odd multiples of $\pi$. This is the case described in the previous Section. 
For the symmetric case of equivalent barriers at the two  N-S boundaries the normal reflection
lifts this degeneracy at the Fermi energy, as can be deduced from 
Eq.~(\ref{level}). We  show below, however, that the lifting of the degeneracy is restricted
in the sense that the amplitude of the giant conductance oscillations remains {\em proportional} to
$N_{\perp}$.

We begin with a qualitative argument and neglect as a first step 
the quantization of the
transverse momentum. Hence we consider $k_{\perp}(n)$  to be a  continuous variable ($k_{\perp}(n)\to
k_{\perp}$).  Within this approximation the spectrum $E_l(k_{\perp})$ and the wave functions $|l,
k_{{\perp}}>$ of a  quasiparticle are characterized by one discrete quantum number $l$ associated
with the longitudinal quantization and by one continuous variable, the transverse wave vector
$k_{\perp}$.  As can be seen from Eq.~(\ref {level})  energy levels are at the Fermi energy
($E=0$) if two conditions are satisfied, viz. % 
\begin{equation}
\phi =\pi (2s_0+1)  \label{phi}
\end{equation}
and
\begin{equation}
\varphi _{+}=2kL =2\pi q_0  \label{phi+}
\end{equation}
where $k$ (we have dropped the subscript $n$) now is a continuous variable; $s_0$ and $q_0$ are
integer numbers. It follows that in the absence of transverse momentum quantization the symmetric
barriers at the N-S boundaries  do not completely remove the degeneracy of the energy level at the
Fermi energy. The extent of the degeneracy depends on the 
number of transverse wavevectors [cf. Eq.~(\ref{phi+})] for which the equation (\ref{level}) is
satisfied. This number is determined by the largest possible value of $q_0$,  which will
be estimated below. 

From its definition one notes that $k=\sqrt{k_F^2-k_{\perp }^2}$ varies between zero and
$k_F$ and hence from (\ref{phi+}) one concludes that $0\leq q_0\leq k_FL/\pi$. This implies that the
maximum value of $q_0$, let's call it  $N_0$, is of order $k_FL \gg 1$.   Therefore, whenever $\phi
=\pi (2s_0+1)$, there is a degenerate energy level at the Fermi level with degeneracy $ \sim N_0$.
The number of states through which an electron can be resonantly transmitted from one reservoir to
the other is even greater, however. This is because the width of the
energy levels  broadened due to the  coupling of the sample to the electron reservoirs is
   $\delta E\sim \epsilon_r\hbar v_F/L$ and  all the quantum states inside 
this range of energy
 resonantly transfer reservoir electrons through the sample. In order to determine the
 number  of states within this energy range we
estimate the total width of the intervals in the $k$-space around points $k=\pi q_0/L$ inside which
wave functions $|l, k_{\perp}>$ of the system correspond to energy levels inside
this range of energy,  $E_l(k_{\perp})\leq \epsilon_r\hbar v_F/L$.
We do so by expanding the cosines in Eq.~(\ref{level}) in a
Taylor series in the small deviations $\delta k$ and $\delta E=\epsilon_r\hbar v_F/L$  near one of
the points where the cosines  are equal to unity [these points are determined by Eqs.~(\ref{phi})
and (\ref{phi+})]. Employing the sum rule $|r_A|^2+|r_N|^2=1$ one can show that the energy levels
are inside the resonant range $E\leq \delta E=\epsilon_r\hbar v_F/L$ if $\delta k\leq
\epsilon_r/|r_N|L$, assuming $|r_N|\gg \epsilon_r$. Multiplication by the number $N_0$ of such
intervals gives the total range of the ``resonant'' momenta as % 
\begin{equation}
\Delta k\sim \frac{\epsilon_r}{|r_N|}k_F,\quad \epsilon_r\ll |r_N|\quad \label{resineq}
\end{equation} 
A similar analysis shows that if $\epsilon_r\geq |r_N|$ all $N_{\perp}$ transverse modes take part
in the resonant transition; the
oscillations  disappear if $|r_A|=[1-|r_N|^2]^{1/2}\ll \epsilon_r$.

Now we go one step further and take the transverse quantization into account. In the limit 
$1/N_{\perp}\ll\epsilon_r\ll 1$ the quantized values of
momentum $k_n=[k_F^2-(n\pi/d)^2]^(1/2)$ are almost evenly distributed between zero and $k_F$. 
Hence it follows that the probability for a transverse mode to be inside the resonant interval
$\Delta k$ is $P=\Delta k/k_F=\epsilon_r/|r_N|$. Therefore the total number of transverse modes
inside the resonant region $\Delta k$ is $N_R\sim N_{\perp}P=N_{\perp}\epsilon_r/|r_N|$. From here
and from Eq.~(\ref{G}) it follows that the maximum
conductance (when electrons are resonantly transmitted through the sample) is
\begin{equation}
G_{max}\propto N_{\perp}\frac{2e^2}{h }\frac{\epsilon_r}{|r_N|}
\quad .
\label{maxcond}
\end{equation}
Analytical calculations presented in Appendix~C, see Eq.~(\ref{Gapp}), show the
conductance of a sample with symmetric N-S boundaries (i.e., boundaries
with equal probabilities of normal reflection)
to be
\begin{equation}
G\simeq N_{\perp}\frac{2e^2}{h}\frac{\epsilon_r^2}{\sqrt{(1+|{ 
r_A}|^2\cos \phi +\epsilon_r^2/2)^2-|{ r_N}|^4}}
\label{GT}
\end{equation}
As is evident from Eq.~(\ref{GT}), the maximum conductance occurs when $\phi =\pi
(2l+1)$, which is when energy
levels line up with the Fermi energy and, therefore, resonant transition of
electrons from one reservoir to the other via the sample takes place. Using
Eq.~(\ref{GT}) it is straightforward to see that the maximal conductance is
\begin{equation}
G_{max}\approx N_{\perp}\frac{2e^2}{h }\frac{\epsilon_r}{\sqrt{
|r_N|^2+\epsilon_r^2/4}}
\label{maxG}
\end{equation}
If $|r_N|\ll \epsilon_r$ we have the giant conductance oscillations
predicted in Ref.~\onlinecite{go}. If $|r_N|\gg \epsilon_r$ the maximal conductance
is determined by Eq.~(\ref{maxcond}). The minimal conductance --- occurring when $\phi =2\pi
l$ ---  when we are off resonance, is
\begin{equation}
G_{min}\approx N_{\perp}\frac{2e^2}{h }\frac{\epsilon_r^2/2}{\sqrt{
|r_A|^2+\epsilon_r^2/4}} 
\; .
 \label{mincond}
\end{equation}
The ratio between the maximal and the minimal conductances is therefore
\begin{equation}
\frac{G_{min}}{G_{max}}\approx \frac{\epsilon_r}2\sqrt{\frac{
|r_N|^2+\epsilon_r^2/4}{|r_A|^2+\epsilon_r^2/4}}
\label{ratio}
\end{equation}
Hence it follows that
\begin{equation}
\frac{G_{min}}{G_{max}}\approx \left\{
\begin{array}{ll}
\epsilon_r^2/4 & |r_N|\ll \epsilon_r \\
\epsilon_r, & |r_N|\sim |r_A| \\
\frac{\epsilon_r|r_N|}{2|r_A|}, & \epsilon_r\ll |r_A|\ll |r_N|\\
1, & |r_A|\ll \epsilon_r
\end{array}
\right.   
\label{rin}
\end{equation}
In a situation when $|r_N|\sim |r_A|$ the
amplitude of the conductance oscillations is greater by a factor $N_{\perp}\epsilon_r\gg 1$ 
than in the absence of the superconducting mirrors. If $|r_N|\leq\epsilon_r$ the amplitude of the
conductance oscillations is $\propto N_{\perp}$.

In the above analysis we considered the case of equivalent boundary potentials, so that
the probabilities of normal reflection are equal at the two N-S interfaces. When these
probabilities are not equal, the energy levels never reach the Fermi energy
and resonant transmission occurs only if the asymmetry is not too large.
Below we analyse the situation of non-equivalent N-S boundary potentials.

\subsection{\label{asymmetric reflection}Normal reflections from non-equivalent N-S boundary
potentials} %
Matching of the wave functions of the electron- and
the hole-like excitations at two non-equivalent N-S boundaries results in a spectral function
of the form
\begin{equation}
Q_n=\cos \varphi _{-}-|{ r_{N}^{(1)}}||{ r_{N}^{(2)}}|\cos \varphi _{+}+|%
{ r_{A}^{(1)}}||{  r_{A}^{(2)}}|\cos \phi \quad   
\label{sfg}
\end{equation}
and the energy levels of the system are determined by solutions to the equation
\begin{equation}
\cos 2mE L/\hbar^2 k_n=|{ r_{N}^{(1)}}||{ r_{N}^{(2)}}|\cos 2k_nL-|{ r_{A}^{(1)}}|
|{ r_{A}^{(2)}}|\cos  \phi \quad   .
\label{sfg1}
\end{equation}
Here $r_{N}^{(1,2)}$ and $r_{A}^{(1,2)}$ 
 are the probability amplitudes for an electron to
be normally and Andreev reflected, respectively, at the left ($1$) and right ($2$)
boundaries; $|r_{N}^{(1,2)}|^2+|r_{A}^{(1,2)}|^2=1$. As follows from Eq.~(\ref{sfg}), if 
 $r_N^{(1)}$ and $r_N^{(2)}$ are different there is an energy gap in the spectrum around the Fermi
energy since the maximal value of the right side of Eq.~(\ref{sfg}) is 
smaller than
unity and hence there is no energy level at the Fermi energy for any $\phi$. For a weak asymmetry
between the boundaries, $\delta r_N=|r_{N1}-r_{N2}|\ll 1$, the maximal value of
the right hand side of Eq.~(\ref{sfg}) differs from unity by an amount
\begin{equation}
\delta M=(\delta r_N)^2\quad 
\label{Mas}
\end{equation}
Hence it follows that resonant transmission of electrons occurs only if $%
\delta r_N\leq \epsilon_r$.
Analytical calculations carried out for the general case in 
Appendix~C 
shows the
conductance to be 
\begin{equation}
G\approx N_{\perp}\frac{2e^2}{h}\frac{\epsilon_r^2}
{\sqrt{(1+|r_{A}^{(1)}||r_{A}^{(2)}|\cos \phi +\epsilon_r^2/2)^2-|r_N^{(1)}|^2|r_N^{(2)}|^2}}
\; .
\label{condass}
\end{equation}
It follows from Eq.~(\ref{condass}) that the maximal conductance is, with $\delta
r_A=|r_{A}^{(1)}-r_{A}^{(2)}|$, %
\begin{equation}
G_{\max }\approx N_{\perp}\frac{2e^2}{h}\frac{\epsilon_r^2}
{\sqrt{\delta r_A^2+\epsilon_r^2(1-|r_{A}^{(1)}||r_{A}^{(2)}|)+\epsilon_r^4/4}}
\end{equation}
Therefore the giant oscillations are of the
same kind as described above if $\delta r_A\leq \epsilon_r$, but the maximal value of the
conductance decreases with increasing $\delta r_A$; when $\delta r_A\gg \epsilon_r$
the maximal value of the conductance is
\begin{equation}
G_{max}\simeq N_{\perp}\frac{2e^2}{h}\frac{\epsilon_r^2}{\delta r_A}\quad 
\label{maxGas}
\end{equation}

	\section{Numerical calculations}\label{numerical calculations}
In the range of parameters where $\epsilon_r$ and hence the coupling between sample
and reservoirs is not small the
approximations used above are not valid and the set of equations 
Eq.~(\ref{match1}) must be solved exactly. In order to find the
largest value of $\epsilon_r$ for which the conductance oscillations
are giant, and to find the dependence of the conductance on
parameters of the system we have resorted to numerical methods. We have
solved the problem for different  coupling strengths (from 20\% to 100\% of the largest value of
$\epsilon_r$ for which the scattering matrix ${\bf \hat S}$ of Eq.~(\ref{Gens}) is still unitary;
see below), for a varying number of transverse modes $N_{\perp}$ (from 5 to 40), and for different
values of the phase difference $\phi$ between the two superconducting condensates (from 0 to
2$\pi$).

To calculate the conductance of our system we use the Lambert formula. The transmission and 
reflection amplitudes are calculated by matching the waves.
Our task is to find the probability amplitudes for $b_1$ and $b_5$ for quasiparticles going
into the reservoirs as functions of parameters of the system and of the amplitudes $a_1$ and $a_5$
of quasiparticles approaching the sample from the reservoirs. One parameter is the number 
of modes $N_{\perp}>1$, which we relate to the the width of the normal
conductors (assuming a 2D system) as  
\begin{equation} 
W=(N_{\perp}+0.5)\lambda_F/2
\; .
\end{equation} %
The matching of amplitudes at the left (1) and right (2) junctions are 
performed using the 
scattering matrix of Appendix~A
\begin{equation}
\left(\begin{array}{c}
b_1\\
b_2\\
a_3
\end{array}\right)
=\hat S_l
\left(\begin{array}{c}
a_1\\
a_2\\
b_3
\end{array}\right)
\end{equation}
\begin{equation}
\left(\begin{array}{c}
b_5\\
\hat u_3^{-1}b_3\\
a_4
\end{array}\right)
=\hat S_r
\left(\begin{array}{c}
a_5\\
\hat u_3a_3\\
b_4
\end{array}\right)
\end{equation}
First we eliminate $a_2$ and $b_4$ by expressing them in terms of $b_2$ and $a_4$,
\begin{equation}\begin{array}{c}
  a_2=\hat u_2\hat R_l\hat u_2b_2=\hat \alpha_lb_2\\
  b_4=\hat u_4\hat R_r\hat u_4a_4=\hat \alpha_ra_4
\end{array}\end{equation}
In the next step we eliminate $a_4$ and $b_2$ and to proceed we first define
\begin{equation}\begin{array}{l}
  \hat \beta_l=\hat \alpha_l(1-\hat s_{22l}\hat \alpha_l)^{-1}\\
  \hat \beta_r=\hat \alpha_r(1-\hat s_{33r}\hat \alpha_r)^{-1}
\end{array}\end{equation}
and then
\begin{equation}\begin{array}{l}
  \hat \gamma_{1l}=\hat s_{31l}+\hat s_{32l}\hat \beta_l \hat s_{21l}\\
  \hat \gamma_{2l}=\hat s_{33l}+\hat s_{32l}\hat \beta_l \hat s_{23l}\\
  \hat \gamma_{1r}=\hat s_{21r}+\hat s_{23r}\hat \beta_r \hat s_{31r}\\
  \hat \gamma_{2r}=\hat s_{22r}+\hat s_{23r}\hat \beta_r \hat s_{32r}\\
\end{array}\end{equation}
Using these quantities we conveniently can find the following expressions for $a_3$ and $b_3$,
\begin{equation}\begin{array}{l}
  a_3=\hat \gamma_{1l}a_1+\hat \gamma_{2l}b_3\\
b_3=(1-\hat u_3\hat \gamma_{2r}\hat u_3\hat \gamma_{2l})^{-1}\hat u_3
(\hat \gamma_{1r}a_5+\hat \gamma_{2r}\hat u_3\hat \gamma_{1l}a_1)
\end{array}\end{equation}
Finally we can calculate
\begin{equation}\begin{array}{l}
b_1=(\hat s_{11l}+\hat s_{12l}\hat \beta_l\hat s_{21l})a_1+(\hat s_{13l}+\hat s_{12l}\hat \beta_l\hat s_{23l})b_3\\
b_5=(\hat s_{11r}+\hat s_{13r}\hat \beta_r\hat s_{31r})a_5+(\hat s_{12r}+\hat s_{13r}\hat \beta_r\hat s_{32r})\hat u_3a_3
\end{array} \; .
\end{equation}
The studied system is symmetric in the sense that the two scattering matrices connecting the
sample and the reservoirs are equal and the probability of normal reflection is
the same for both superconducting mirrors. These symmetries makes further simplifications possible.

According to the discussion in section \ref{coupling} (see Ref.~\onlinecite{note_order}) the 
scattering matrix 
${\bf \hat S}$ in Eq. (\ref{Gens}) can be taken to be a random matrix. For our numerical
calculations we determine it as described in Appendix~A.

\begin{figure}%%%[hbt]\hskip 1cm\hbox
\centerline{\psfig{figure=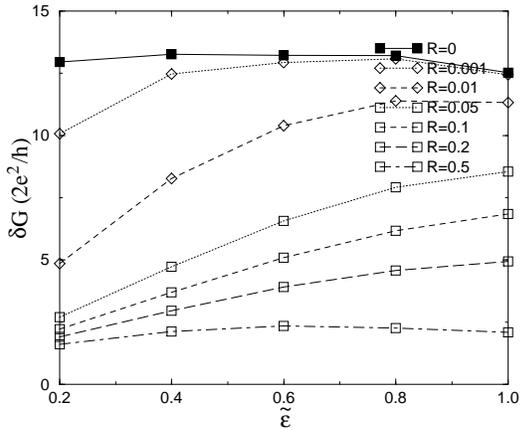,width={8 cm}}}
%%%\vspace{0.5cm}
\caption{\protect{\label{fig2a}}
Difference $\delta G$ between
maxima and minima in the conductance as a function of the parameter 
$\tilde\epsilon$,
which characterizes the junction scattering matrices (realized by first
method mentioned in text). 
Results are plotted for varying probabilities
$R=|r_N|^2$ for normal reflection at the superconducting mirrors. 
The number of transverse modes are $N_{\perp}=40$. The results agree with the
weak coupling limit calculated analytically, see Eq.~(58).
}
%%%\label{fig2a}}\vskip .5cm
\end{figure}

\begin{figure}%%%[hbt]\hskip 1cm\hbox{
%%%%\vspace{0.5cm}
\centerline{\psfig{figure=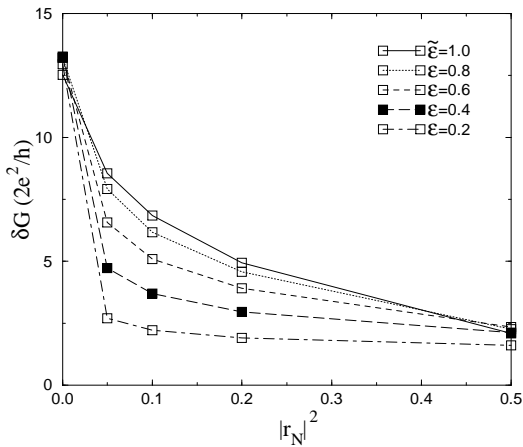,width={8 cm}}}
\caption{\protect{\label{fig4a}}
Difference $\delta G$ between
maxima and minima in the conductance as a function of the
probability $R=|r_N|^2$ for normal reflection at the superconducting mirrors
(same data as in Fig.~6).
The results agree with the
weak coupling limit calculated analytically, see Eq.~(58).
}
%%%\label{fig4a}}\vskip .5cm
\end{figure}

The scattering matrix describing coupling and mode mixing at the junctions have been realized in
two  different ways. First by assigning random numbers to its elements. Here a critical value 
$\epsilon_c$ of
the coupling strength $\epsilon_r$ was found in the sense that the scattering matrix was
non-unitary for $\epsilon_r>\epsilon_c$. We find it convenient to define a new parameter
$\tilde\epsilon\equiv \epsilon_r/\epsilon_c$, which can be varied between 0 and 1. The results from
these calculations are shown in Figs.~\ref{fig2a}-\ref{fig4a}. Every point is an average of 10
realizations of the random scattering matrix. The spread in conductance was $4e^2/h$ when normal
reflection was absent at the N-S interfaces and $2e^2/h$ when the normal reflection probability was
at its highest studied value. The position of the peak was not seen to change for different
realizations. The critical value of the coupling was in this case determined by the highest
eigenvalue. This gave as a result that only some modes where strongly coupled in the limit of high
$\tilde\epsilon$. 


\begin{figure}%%%[hbt]\hskip 1cm\hbox
\centerline{\psfig{figure=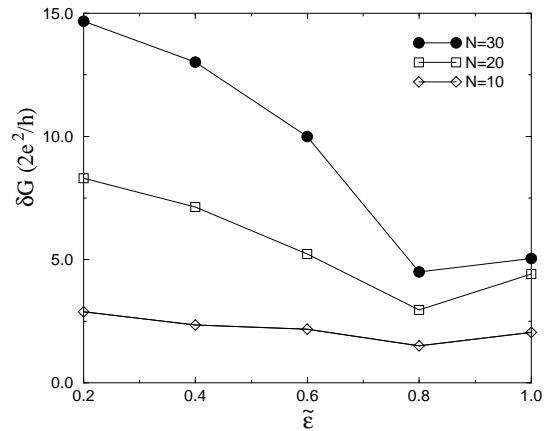,width={8 cm}}}
%%%\vspace{0.5cm}
\caption{\protect{\label{fig0}}
Difference $\delta G$
between maxima and minima in conductance  as a function of coupling
$\tilde\epsilon$ (scattering matrix realized by second method, see text). 
The probability for normal reflection $|r_N|^2=0$. The results
agree with the analytical results in the weak and intermediate range of
coupling where the resonant peak is proportional to the number of channels.
}
%%%\label{fig0}}\vskip .5cm
\end{figure}

\begin{figure}%%%[hbt]\hskip 1cm\hbox{
\centerline{\psfig{figure=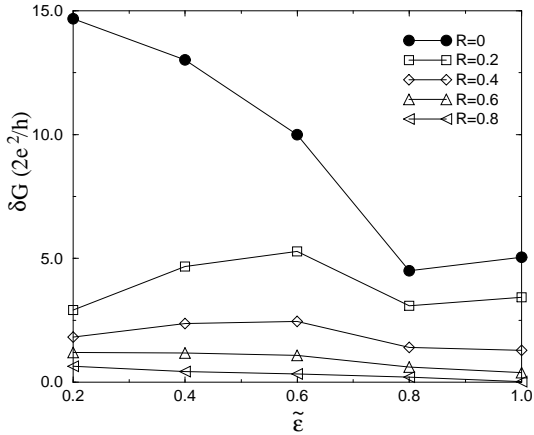,width={8 cm}}}
%%%\vspace{0.5cm}
\caption{\protect{\label{fig30}}
Difference $\delta G$
between maxima and minima in conductance  as a function of coupling
$\tilde\epsilon$. The number of open transverse modes are
$N_{\perp}=30$. The results agree with Eq.~(58) for weak coupling.
For strong coupling the giant effect vanishes in a 2D sample according to the 
discussion in Section~I.
}
%%%\label{fig30}}\vskip .5cm
\end{figure}

\begin{figure}%%%[hbt]\hskip 1cm\hbox{
\centerline{\psfig{figure=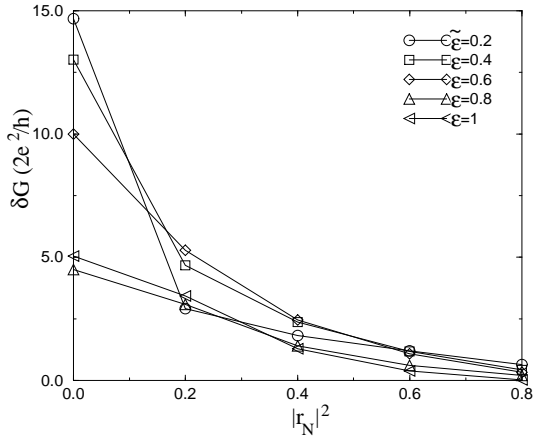,width={8 cm}}}
%%%\vspace{0.5cm}
\caption{\protect{\label{figR}}
Difference $\delta G$
between maxima and minima in conductance as a function
of normal reflection probability $|r_N|^2$
(same data as in Fig.~9). The number of open transverse modes
is $N_{\perp}=30$, results for different strength of the coupling are shown. 
The results agree with analytical calculations.
The $|r_N|^2$-dependence agress with Eq.~(58). 
}
%%%\label{figR}}\vskip .5cm
\end{figure}

\begin{figure}%%%[hbt]\hskip 1cm\hbox{
\centerline{\psfig{figure=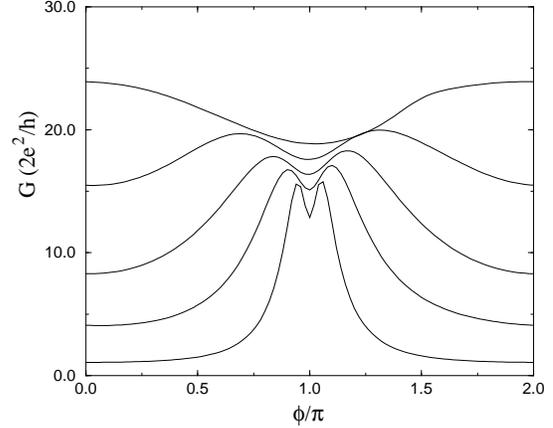,width={8 cm}}}
%%%\vspace{0.5cm}
\caption{\protect{\label{fig.peaks}}
The resonance peaks at zero probability for normal reflection at
the N-S boundaries. The results are from numerical calculations with
$N_{\perp}=30$ and for different strength of coupling 
$\tilde\epsilon=[0.2, 0.4,
0.6, 0.8, 1.0]$ where the most narrow peak is for weakest coupling
$\tilde\epsilon=0.2$. Note that the amplitude of oscillation is much
larger than the conductance quantum
even for $\tilde\epsilon=1$. This is
 because quasiparticle waves may pass the junction in our
realization of the scattering matrix even if $\tilde\epsilon=1$.
}
%%%\label{fig.peaks}}\vskip .5cm
\end{figure}

The second type of realization of the scattering matrix 
was done with B\"uttiker
matrices \cite{Buttiker84} describing a coupling mode by mode 
between sample and reservoirs.
In this case
an additional unitary matrix was used, which only mixed the modes, 
see Appendix~A. Both matrices were parametrized by the coupling 
parameter  $\tilde\epsilon$. The result
of these calculations are shown in Figs.~\ref{fig0}-\ref{figR}. 
The only parameter to be changed
in order to get different realizations of the random scattering 
matrix was an angle $\varphi_{ii}$
which only changed the position of the resonant peak. For zero 
angle the shape of the peak is seen
in Fig.~\ref{fig.peaks}. The number of open modes are in this 
realization equal to the size of
the matrix as all eigenvalues have an amplitude of unity. 

The main result from the analytical calculations to be compared with the numerical results is 
$G_{max}-G_{min}$. This is in general approximately equal to $G_{max}$. From equation
(\ref{maxcond}) we get 
\begin{equation}
G_{max}\propto N_{\perp}\frac{2e^2}{h}\frac{\epsilon_r}{|r_N|}
\quad 
\end{equation}
which agrees with numerical results when $\epsilon_r<|r_N|$. 

The first realization of the random scattering matrix has been found to describe the weak coupling
case as the observed peaks  were narrow even for $\tilde\epsilon=1$. The second realization with a
separate matrix mixing modes gave the possibility to study weak and intermediate coupling and the
amplitude of oscillations were seen to diminish when $\tilde\epsilon$ was increased, 
see Fig.~\ref{fig.peaks}. %


	\section{Giant conductance oscillations for a diffusive normal sample - Feynman path integral 
approach.}\label{Feynman}
In this Section we want to study the conductance oscillations by considering 
the probability amplitude for transmission and
reflection of  electrons and holes  between the 
reservoirs via an  S-N-S system the diffusive transport regime (see Fig.~\ref{fig22}) as a sum of
Feynman paths. \cite{Feynman} As we will show below,  one does not actually
need to do any complicated summations to find this probability amplitude, because  for electron
energies below the Thouless energy $E_c$ (or equivalently for temperatures below the Thouless
temperature $T_c =E_c/k_B$) the hole exactly retraces the
electron diffusive path after Andreev reflection. It follows that the phase gain of the electron
along any resonant path between the N-S boundaries (see Fig.~\ref{fig22}) is  compensated  by the
hole phase gain along the same path. Therefore the phase gain is determined only by the phases
imposed on the quasiparticles by the superconductors  when a trajectory
encounters the  N-S boundaries.   As a result the amplitude does not depend on either the
form, the length of the diffusive  path between the superconductors  or the configuration of
impurities (which means there is no need to perform any ensemble averaging of the
conductance).  The dependence of the resonant probability amplitude on the phase difference between
the superconductors and on the scattering amplitudes at the barriers is easily found by calculating 
the number of reflections at the N-S boundaries and the number of backscattering events at the
barriers.  The conductance is equal to the probability of transmission (the modulus squared of
the probability amplitude) multiplied by the number of different classical  resonant paths
(more strictly, by the number of tubes of width $\sim \lambda_F$ around these paths
\cite{Feynman}) starting out from a reservoir lead; a number that can be straightforwardly
estimated.  We emphasize again that since the conductance associated with resonant
transmission and reflection does not depend on the impurity configuration there is no need to 
average it with respect to the impurity positions.  
 
We start by deriving an equation that connects the Feynman path integrals for
electrons and holes. To do this we shall need the boundary conditions at an N-S boundary for 
the relevant Green's functions.

The probability amplitude $K^{(e,h)}({\bf r}, {\bf r}^{\prime}; t-t^{\prime})$ for an electron 
(hole)  to propagate  from point ${\bf r}$ at time $t$ to point ${\bf r}^{\prime}$ at time
$t^{\prime}$ is given by the time dependent Green's function satisfying the following equation 
\begin{equation}
\left(\mp i\hbar \partial/\partial t +\hat{H} \right) K^{(e,h)}({\bf r}, 
{\bf r}^{\prime};t-t^{\prime})=\delta ({\bf r}- {\bf r}^{\prime})\delta(t-t^{\prime}) 
\; .\label{time}
\end{equation}
Here the plus (minus) sign is for electrons (holes). The initial condition is
\begin{equation}
 K^{(e,h)}({\bf r}, {\bf r}^{\prime};t-t^{\prime})=0 \;\;\mbox{for}\;\; t-t^{\prime}<0
\label{initial}\end{equation}
and$\hat{H}$ in Eq.~(\ref{time})  is the Hamiltonian describing a metal in the diffusive transport
regime: %
\begin{equation}
\hat{H}=-(\hbar/2m) \nabla^2 + V_{imp}({\bf r})-\epsilon_F.
\label{Ham}\end{equation}
The potential $V_{imp}$ is
\begin{equation}
 V_{imp}({\bf r})=\sum _{j}v({\bf r}-{\bf R}_j),
\label{imp}\end{equation}
and $v({\bf r}-{\bf R}_j)$ is the potential of an impurity at point ${\bf R}_j$.

In order to derive the boundary conditions we observe that the time
Fourier transform of $K^{(e,h)}({\bf r}, {\bf r}^{\prime};t-t^{\prime})$ for the case 
of electrons 
satisfies the equation
\begin{equation}
(\hat{H} - E-i\eta) K^{(e)}_{E}({\bf r}, {\bf r}^{\prime})=
\delta ({\bf r}- {\bf r}^{\prime})
\label{gel}\end{equation}
while for the hole case the equation is
\begin{equation}
(\hat{H}+E+i\eta)K^{(h)}_{E}({\bf r}, {\bf r}^{\prime})=\delta ({\bf r}- {\bf r}^{\prime})
\label{gh}\end{equation}
($\eta$ is a small positive constant).
At the N-S boundaries the Green's functions
$K^{(e,h)}_{E}({\bf r}, {\bf r}^{\prime})$ are connected with each other by
the Andreev reflection condition for a fixed energy:
\begin{equation}
K^{(h)}_{-E}({\bf r}^{(1,2)}, {\bf r}^{\prime})=e^{i(\Phi_{1,2}+\Psi_E)}
K^{(e)}_{E}({\bf r}^{(1,2)}, {\bf r}^{\prime})
\; .
\label{boundE}\end{equation}
Here ${\bf r}^{(1,2)}$ and $\Phi_{1,2}$ are the coordinate and the phase of the gap function at 
the first (second) N-S boundary, respectivly, and 
${\rm e}^{i\Psi_E}=|\Delta|/(E-i\sqrt{|\Delta|^2-E^2}) $, where $|\Delta|$ is the magnitude
of the gap.
Now, an inverse Fourier transformation of Eq.~(\ref{boundE}) results in the following
relation,
\begin{eqnarray}
\nonumber
K^{(h)}({\bf r}^{(1,2)},{\bf r}^{\prime} ; \tau )&=& 
{\rm e}^{i\phi_{1,2}}\int_{-0}^{\infty} d\tau^{\prime} K_E^{(e)}({\bf r}^{(1,2)},{\bf r}^{\prime}
;  \tau^{\prime} ) \\
&&\times\int_{-\infty}^{\infty}dE {\rm e}^{i\Psi_E}e^{i(\tau^{\prime}-\tau)E/\hbar}
\label{Four}
\end{eqnarray}
where $\tau =t-t^{\prime}$. 
We are interested in the case when the characteristic time of transmission $t_2-t_1$ is of the 
order of the time, $L^2/D$, it takes to diffuse the length $L$ of the sample. Since this time can
be expressed in terms of the Thouless energy as $\hbar/E_c$, the characteristic time
difference $|\tau-\tau^{\prime}|$ in the last integral of Eq.~(\ref{Four}) is of the order of
$L^2/D=\hbar/E_{c}$. Therefore \cite{fot2} in the last integral of Eq.~(\ref{Four})  the main
contribution is from energies inside an energy interval of order $E_{c}\ll |\Delta|$.
In this interval $\Psi_E \approx \pi/2$ and hence the second integral in Eq.~(\ref{Four}) is
a Dirac $\delta$-function. Therefore the boundary condition for the time-dependent electron- and
hole Green's functions at the N-S boundaries is  
\begin{equation}
 K^{(h)}({\bf r}^{(1,2)}, {\bf r}^{\prime}; \tau)=
e^{i(\phi_{1,2}+\pi/2)} K^{(e)}({\bf r}^{(1,2)}, {\bf r}^{\prime}; \tau) 
\label{timebound} 
\end{equation}

According to the Feynman approach \cite{Feynman} a probability amplitude
can be written as a path integral. Here we are interested in the following probability amplitude,
\begin{equation}
K({\bf r}_1,t_1;{\bf r}_2,t_2) =
\int_{({\bf r}_1,t_1)}^{({\bf r}_2,t_2)} e^{iS \{{\bf r}(t)\}/\hbar}D\{{\bf r}_e(t)\}D\{{\bf r}_h(t)\} \label{Green}
\end{equation}
 This expression sums over all possible  paths of an electron  which
start from a point ${\bf r}_1$ in the first reservoir at time $t_1$  and end at a  point ${\bf
r}_2$ at time $t_2$ in either the first or second reservoir and of all possible hole paths
that appear due to Andreev reflections at the N-S boundaries. For any path the classical action is
\begin{equation}
 S=\int_{t_1}^{t_2}{\cal L} dt+\Psi_A
\; ,
\label{action}\end{equation}
where the phase $\Psi_A$ will be discussed below. The Lagrangian ${\cal L}$
for those sections of the
path where the particle moves as an electron is %
\begin{equation}
{\cal L}=m\dot{{\bf r}}^2_{e}-V\left({\bf r}_{e}(t)\right)
\label{Lagr.e}
\end{equation}
For those sections of the path where the particle moves as a hole we have
\begin{equation}
{\cal L}=-m\dot{{\bf r}}^2_{h}+V\left({\bf r}_{h}(t)\right) \; .
\label{Lagr.h}\end{equation}
Here $m$ is the electron mass, ${\bf r}_{e}$ and ${\bf r}_{h}$ are the electron  and hole 
coordinates, respectively, the potential is $V=V_0+V_{imp}$; $V_0$ describes the barriers between
the sample and the leads to the reservoirs [$V_{imp}$ is defined in Eq.~(\ref{imp})]. While
performing the integration in Eq.~(\ref{Green}) one has to use the N-S boundary conditions 
for electron and hole trajectories given by Eq.~(\ref{timebound}). The boundary conditions
results  in an additional term $\Psi_A$, which depends on the macroscopic phases of the
superconductors: %
\begin{eqnarray}
\nonumber
\Psi_A=&&(\pi/2 +\phi_1)P^{(1)}_e+(\pi/2 + \phi_2)P^{(2)}_e\\
&&+
(\pi/2 -\phi_1)P^{(1)}_h+(\pi/2-\phi_2)P^{(2)}_h
\label{Psi}
\end{eqnarray}
In this expression $P^{(1)}_{(e,h)}$ and $P^{(2)}_{(e,h)}$ count how many electron-hole
(hole-electron)  transformations that has occurred at the N-S boundaries for a certain trajectory.

Transport properties of a diffusive system are  usually calculated  in the semiclassical 
approximation, which implies (for instance) that the 
cross section for impurity scattering is larger than  $\lambda_B^2$. We adopt this point of view
when we now proceed to calculate the functional integral in Eq. (\ref{Green}). 
This means that the method of steepest descent is useful for performing the integration in
Eq.~(\ref{Green}), and  hence classical trajectories that minimize the action Eq.~(\ref{action})
contribute to the integral. \cite{Feynman} 

In order to get the probability amplitude for transmission of an electron having fixed energy 
$E$ from point ${\bf r}_1$ of  reservoir 1 to point ${\bf r}_2$ of reservoir 1 or 2 at all times
one has to sum all the relevant amplitudes at all times, that is to integrate the amlitude $K$
(multiplied by a factor $\exp(iE \tau/\hbar)$) with respect to time $\tau=t_2-t_1$. From this
we conclude that the probability amplitude $A(E)$ for transmission (or reflection) is equal to  %
\begin{equation} 
A(E)=\sum_{\{S\}}R(S)exp\left[\int_{({\bf r}_1)}^{({\bf r}_2)}p_S(s)ds/\hbar+
\Psi_A(S)\right] 
\; ,
\label{ampl}
\end{equation}
where summation is with respect to  classical trajectories $S$ that start at point ${\bf r}_1$ and 
end at point ${\bf r}_2$ in reservoirs along which an electron with energy $E$ reaches reservoir 2
(since an electron or a hole) starting from reservoir 1, or is reflected back into  reservoir 1 (as an
electron or a hole); $p_S(s)$ is the classical momentum as a function of the coordinate $s$ along
trajectory $S$, being equal to the electron momentum $p_e$ and hole momentum $p_h$ at the electron-
and hole sections of the trajectory $S$, respectively. $R(S)$ is a product of the probability
amplitudes of reflection and transition at the barriers between the sample and the reservoirs that
occur  for the electron and hole along  path $S$. $\Psi_A(S)$ is the phase gained along the
path $S$ by Andreev reflections at the N-S boundaries. When counting the
number of the trajectories one has to take into account the fact that the  trajectories  has to be
considered as tubes with a width of the order of the de Broglie wavelength $\lambda_{dB}=h/p_F$
(see above).

Now we can calculate the probability amplitudes for an electron at the Fermi level with energy
 $E=0$ from one reservoir to be reflected as a hole back to the same reservoir and
to be transmitted as an electron to the other reservoir via the diffusive normal metal part of
the sample.  
It is crucial for the calculation that at $E=0$  under Andreev reflection the hole and electron  
momenta are equal but their velocities have equal magnitude but opposite signs. This means that the
classical trajectories of the electron and hole that end and start at the same  points at the N-S
boundaries exactly repeat each other in both ballistic and  diffusive samples (as the
classical trajectory is uniquely determined by the starting point and the velocity of the
particle). Hence it follows that for any classical trajectory with Andreev reflections, at $E=0$ 
the total classical action $\int_{({\bf r}_1)}^{({\bf r}_2)}p_S(s)ds/\hbar$ (which is the sum of
the electron and hole actions) is equal to zero as the electron and hole momenta are equal
($p_e=p_h$) and the integrations are along the same trajectory but in opposite directions.
Therefore the phase gain along such trajectories [see Eq.~(\ref{ampl})] does not depend on either
their form, the length of the diffusion path or on the configuration of impurities. For
resonant transmission the summation in Eq.~(\ref{ampl}) with respect to the ampencondlitude scatterings at
the junctions is easily carried out in the case of low transparency of the barriers at the
junctions.  

The conductance of a hybrid sample containing both normal metal and superconductors, the
normal conductance is determined \cite{fot3} by the Landauer-Lambert  formula 
(\ref{gencond}),
which for a symmetric system reduces to \cite{Lambert1}  
\begin{equation}
 G=\frac{2e^2}{h}(T_0+R_A)
\label{lamb}
\end{equation} 
The probabilities $T_0$ and $R_A$ were defined above 
[cf. Eq.~(\ref{gencond})]. It is
important to note that trajectories which connect the two reservoirs, necessarily have
a different number of electron- and hole sections, while for  trajectories which start and end in
the same reservoir these numbers are equal. This is a crucial circumstance when one sums amplitudes 
in order to get the total transmission amplitude and implies that there is no complete compensation
of the electron and hole phase gains along those  trajectories which contribute to the transmission
amplitude of quasiparticles. As a result destructive interference suppresses  the transmission
amplitude, and the main contribution to the conductance is from those trajectories along which the
electron is reflected back into the same reservoir as a hole. This is the channel to be discussed
below. A classical path corresponding to this type of reflection is shown in
Fig.~\ref{fig22}.  

After passing the beam splitter at junction $A$ (that is after tunneling through the barrier of 
this junction) the classical diffusive electron trajectory can first encounter either the left N-S
boundary (clockwise motion) or the right N-S boundary (counter-clockwise motion). Adding
the amplitudes of clockwise and counter-clockwise trajectories  (they form a geometric
series) and expanding the amplitudes  in $\delta\phi=\phi-\pi\ll 1$  one finds the total
probability  for an electron being reflected back into the same reservoir as a hole to be 
\begin{equation}
 R_A=\frac{\gamma^2\delta\phi^2}{(\gamma^2+\delta\phi^2)^2}
\; .
\label{holetotal}
\end{equation} %
Here $\gamma\ll 1$ is the probability to pass through the barrier at the junction.

From Eq. (\ref{holetotal}) it follows that the  electron-hole backscattering amplitude is 
zero if $\phi=\pi$. This is due to the interference between the clockwise and
counter-clockwise  quasiparticle trajectories (in the sense discussed above) and can be 
explained as follows.
The amplitude of the electron-hole backscattering can be represented as a sum of 
contributions arising from trajectories with different number of Andreev reflections at the
superconductors. The ratio between successive terms in this geometric series is equal to the
amplitude of one Andreev reflection at each of the two N-S boundaries. Therefore it depends on the
phase difference between the superconductors and becomes equal to one at resonance, when
trajectories with very large number of Andreev reflections give the same contribution as the ones
containing only few Andreev events. This is of course the reason why a resonance in
the electron-hole backscattering channel occurs. In addition all terms in the series
will be multiplied by a factor ${\rm e}^{i\phi_s}$ where $s$  labels the N-S boundary
from which the electron first is Andreev reflected. In our notation $s=1$ for clockwise
trajectories and $s=2$ for counter-clockwise trajectories. An important consequence of the
existence of these multipliers is that on resonance, when $(\phi_1-\phi_2=\pi)$, the ratio of this
extra exponents for clockwise and counterclockwise trajectories is equal to $-1$ and the
resonant contributions from clockwise and counterclockwise trajectories to the amplitude  for
electron-hole backscattering cancel each other. 
A visible manifestation of this cancellation effect is a  splitting of the resonant conductance peak
near $\phi=\pi$ so that $G(\phi=\pi)=0$. \cite{fot4}
 
If there are Schottky barriers at the N-S boundaries additional multipliers appear in the
amplitudes for clockwise  and counterclockwise trajectories. These are $r_A^{(s)}{\rm e}^{i\psi_s}$
($r_A^{(s)}$ is the amplitude of Andreev reflection at the $s$:th N-S boundary ($s=1, 2$)). 

In the case of non-equivalent barriers, $r_A^{(1)}\neq r_A^{(2)}$, there is
no compensation of the  clockwise and counterclockwise contributions as is the case when
$r_A^{(1)}= r_A^{(2)}$=1. In fact, if $r_A^{(1)}\ll r_A^{(2)}$ the splitting of the resonant
peak disappears.

\begin{figure}%%%[hbt]\hskip 1cm\hbox{
\centerline{\psfig{figure=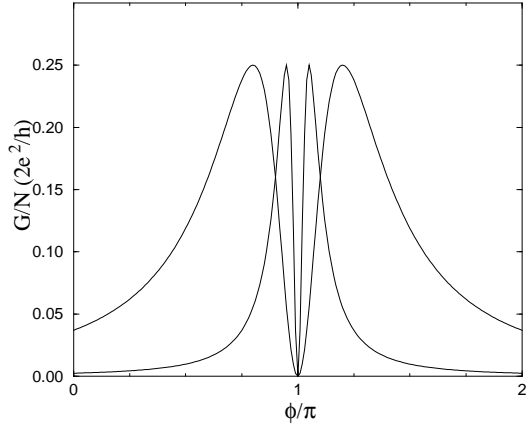,width={8 cm}}}
%%%\vspace{0.5cm}
\caption{\protect{\label{figsplit}}
The resonant conductance peaks are split due to interference between clockwise
and counter-clockwise quasiparticle motion along the trajectories that are
associated with the formation of Andreev levels. Cf. Eq.~(76); here $\gamma$
is 0.05 and 0.2.
}
%%\label{figsplit}}\vskip .5cm
\end{figure} %

In the semiclassical approximation the total number of electrons that  contribute to the 
resonant phase-sensitive conductance is equal to the number $N_{\perp}$ of semiclassical
tubes of  diameter $\lambda_F$ that cover the cross-section of a lead between
the reservoir  and the diffusive sample (assuming the lead has a smaller cross section than the N-S
boundaries). Hence from  the Lambert formula (\ref{lamb}) and
Eq.~(\ref{holetotal}) it follows  that in the semiclassical approximation the phase-sensitive
resonant conductance for a diffusive sample  is equal to 
\begin{equation}
G=(2e^2/h)N_{\perp}\frac{\gamma^2\delta\phi^2}{(\gamma^2+\delta\phi^2)^2}
\label{cond}\end{equation}

Equation~(\ref{cond}) implies that the resonant conductance peaks are split
in  such a way that the conductance goes to zero when  $\phi$ is an odd multiple of
$\pi$ (see Fig.~\ref{figsplit}). This splitting appears due to the interference between the 
clockwise
and counte-clockwise motion of the particles inside the normal sample when electrons are
reflected as holes back to the same reservoir (see above).

The above calculations give a qualitative explanation to  analytical  and numerical results 
for the diffusive case presented in 
Refs.~\onlinecite{Naz,Zaitsev,Lambert,Claughton,Allsopp,Claughton Lambert,Volkov Zaitsev} 
if the results are obtained for low barrier transparency of the junctions between the sample and 
the reservoirs. 

It should be noted that for the geometry considered in some of the papers cited above, where there 
is only one reservoir present, the conductance is determined only by the probability for an
electron to be scattered back into the reservoir as a hole. Therefore  the conductance is
determined by the same equation (\ref{cond}) and hence must also be  equal to zero at
$\phi$ equal to odd multiples of $\pi$ for the equivalentN-S barrier case. The
splitting must disappear for nonequivalent barriers at the N-S boundaries  (see, e.g., Fig.~6
and Fig.~7 in Ref.~\onlinecite{Volkov Zaitsev};  a decrease of the barrier transparency
at the junction between the sample and the reservoir results in the conductance peaks being close
to those  shown in Fig.~(\ref{figsplit}) of this paper). %

We conclude this Section by using the Feynman path integral approach to qualitatively consider the
temperature  dependence of the oscillating part of the conductance for the diffusive case and for
temperatures above the Thouless temperature $E_c/k_B$.
If the energy of an electron-hole excitation is not equal to zero there is no 
exact compensation
of the phases gained along the electron- and hole portions of the paths connected by Andreev
reflections. In this case the phase of the transmission amplitude $A$ depends on the lengths of the
electron-hole  paths, which in their turn depend on the starting points inside
the lead between sample and reservoir. The conductance of the system is a sum of absolute
squares of
amplitudes corresponding to trajectories with different starting points in the lead for the
classical paths. On the
other hand  classical paths starting from points separated by a distance greater than
$\lambda_F$, meet different random sets of impurities \cite{fot5}.
As a result their path lengths have random values. Hence it follows that one can change the
summation over starting points to a summation over path lengths when calculating the
averaged conductance.  For this purpose we assume a Gaussian length distribution of the 
diffusive paths that start at one N-S boundary and end up at the other N-S boundary (we choose the
Gaussian form of the distribution function  as an example; simple considerations show that choosing
a more general distribution function  only results in additional factors of order
unity), 
\begin{equation} 
F({\cal L})=\frac{1}{\sqrt{\pi}{\cal L}_0}\exp\left(-\frac{{(\cal L}-{\cal
L}_0)^2}{{\cal L}_0^2}\right).   
\label{distrib} 
\end{equation}  Here ${\cal L}_0$ is the average length of
the paths  (${\cal L}_0 = L^2 v_F/D$) .  By averaging  the conductance  at a fixed  energy over
random path lengths described by the distribution function (\ref{distrib}) one easily finds a
cut-off factor  of order $\exp \left(-[E/E_{c}]^2/4\right)$  appearing in the interference terms of the
conductance. Therefore,  destructive interference sets in at $E\gg E_{c}$ (this well known
fact justifies the form of  the distribution function assumed above). The conductance
oscillations caused by interference only occur for energies below or of
the order of $E_{c}$. As a result, at temperatures $T\gg E_{c}/k_B$ the amplitude of the conductance
oscillations decreases with the temperature as $E_{c}/T$ in agreement with 
Refs.~\onlinecite{Spivak,Nazarov Stoof,Volkov Lambert}.  

In order to find  the temperature dependence of the giant conductance oscillations
discussed here when $T\gg E_{c}/k_B$ and for low  junction barrier transparencies 
 we sum over the paths contributing to the resonance effect at a certain energy $E$,
average the conductance over the path lengthsusing the distribution function
(\ref{distrib})   and integrate  over energy taking the factor $(-\partial f_0/\partial E)$ 
properly into account ($f_0$
is the Fermi function). As a result we find that the oscillating part of the conductance caused by
the resonant effect is 
\begin{equation} 
\delta G_{res}= N_{\perp}\frac{e^2}{h}
\frac{\epsilon_r E_{c}}{k_B T}g(\phi) 
\label{temposc} 
\end{equation} 
Here $\epsilon_r$ is the
transparency of the barrier at the junction, $g(\phi)$ is a $2\pi$-periodic temperature
independent function with
an amplitude of order unity. 

The physical reason for the result (\ref{temposc}) is that the 
position of the  
resonant energy peak is tuned by the superconducting phase difference 
$\phi$. With a change of
$\phi$ it can be inside or outside the energy interval of  order $E_{c}$ 
associated with the 
conductance oscillations. As the width of the resonant peak is  
$\delta E\sim\epsilon_r E_{c}$ the main contribution
to the conductance oscillations comes from the energy interval $E\sim \epsilon_r E_{c}$, and hence
the relative number of quasuparticles contributing to the oscillations is $\epsilon_r E_{c}/k_B
T$. This is why this factor appears in Eq.~(\ref{temposc}).  

	\section{Conclusions}\label{Conclusions}

In this paper we have presented a more thorough discussion 
than in a previous short communication \cite{go} of giant conductance oscillations in 
hybrid mesoscopic systems of the Andreev interferometer type, i.e. S-N-S structures where the 
N-part is connected to normal electron 
reservoirs. 
In Ref.~\onlinecite{go} giant conductance oscillations were predicted for a ballistic normal sample 
 when transverse mode mixing was absent. The origin of this effect is a degeneracy 
(``bunching" effect) of the 
Andreev energy levels at the
Fermi energy. This degeneracy of the Andreev spectrum arises due to an equality of the
longitudinal momenta of Fermi energy electrons and -holes undergoing Andreev reflection. Any
process that violates this equality lifts the degeneracy and, therefore, decreases the
amplitude of the conductance oscillations.

In this paper we considered the effect of giant conductance oscillations taking into 
account transverse mode mixing at the junctions between the normal part of the sample and the
reservoirs. We also considered normal reflection in addition to Andreev reflection at the
N-S boundaries, and scattering of electrons and holes by impurities inside the normal sample.

Normal reflection of quasiparticles at N-S boundaries decreases the propability of Andreev 
reflection and as a consequence also decreases the amplitude of the conductance
oscillations. We have shown that the probability amplitude for the oscillations is giant
(that is proportional to the number of transverse modes $N_{\perp}$) until the amplitude of
the normal reflection is smaller or of the same order as the transparency $|\epsilon_r|$ of
the barriers at the junctions.

We have also shown that giant oscillations survive in a diffusive sample at temperatures 
much lower than the Thouless temperature. This is because after the electron-hole
transformation associated with an Andreev reflection the electron and the hole move along the 
same classical
diffusive trajectory in opposite directions but with equal momenta. As a result the phase
gain of the electron and the hole along this diffusive path compensate each other. The
probability amplitude for transmission through the sample does not depend on the form or the
length of the diffusive path, but only on the phase difference between the superconductors
(i.e., there is no destructive interference). The number of all possible different
semiclassical paths is $S/\lambda_B^2=N_{\perp}$, where $S$ is the cross-section area, as
each path has a width of the order of the de Broglie wave length $\lambda_B$. Therefore the
amplitude of the conductance oscillations in the diffusive case remains giant and proportional
to the number of transverse modes $N_{\perp}$ as for ballistic samples. The above
qualitative picture agrees with analytical calculations for the diffusive case  by Beenakker
{\em et al.} \cite{Beenakker}, Zaitsev \cite{Zaitsev}, Allsopp {\em et al.} \cite{Allsopp}, 
Volkov and Zaitsev \cite{Volkov Zaitsev}, and Claughton {\em et al.} \cite{Claughton96b}. 

The presence of potential barriers at the junctions between the sample and the normal electron 
reservoirs is most crucial for the giant oscillations to exist.  In the weak coupling case
($\epsilon_r \ll 1$) we have shown analytically and numerically that the amplitude of the
conductance oscillations is independent of the barrier transparency and proportional to the
number of transverse modes $N_{\perp}$. When the transparency of the barriers is increased,
our numerical calculations show that the amplitude decreases. In the absence of barriers at
the junctions the amplitude becomes zero. The latter result agrees with the sum rule in 
Ref.~\onlinecite{Allsopp} according to which the conductance is equal to the number of transverse
modes --- in the absence of barriers --- times the conductance quantum and does not 
depend on the superconducting phase
difference. This can be qualitatively understood since an electron (hole) coming from the
reservoir after being first Andreev reflected at one N-S boundary as a hole (electron) then
at zero temperature returns to the reservoir by retracing the path of the incoming particle
without reaching the second N-S boundary. 

Recently Nazarov and Stoof \cite{Nazarov Stoof},  and Volkov {\em et al.} 
\cite{Volkov Lambert} proposed a new mechanism for conductance oscillations in diffusive
samples that is effective if the temperature is close to the Thouless temperature 
(thermal effect).
They used the dependence of the diffusion coefficient on the quasiparticle energy and found
conductance oscillations with the superconducting phase difference in the absence of
barriers. The amplitude of the oscillations was found to reach its highest value
 at the Thouless energy.
We propose that this effect can be qualitatively understood if one takes into account
the fact that under Andreev transformation at an N-S boundary there is a finite angle between the
trajectories of the incident- and reflected particle, which is proportional to their 
excitation energy.
Simple estimations show that near the Thouless temperature the classical trajectories of the
electron and the hole can be separated by a distance of the order of the de Broglie
wavelength (that is the width of the semiclassical trajectories) before the particle leaves
the normal diffusive part of the sample for a reservoir. When this happens the trajectories
meet different sets of impurities and can diffuse along very different paths inside the
sample. This permits the quasi-particles to encounter also the other N-S boundary and 
undergo Andreev
reflection there before leaving the sample. Therefore the conductance starts to depend on
the superconducting phase difference and conductance oscillations arise. When the
temperature is higher than the Thouless temperature, the phase gains of the quasi-particles
along the semiclassical paths are not compensated and are much larger than unity; the
destructive interference kills the thermal effect, in agreement with the results of the
papers cited above. 

When the transparency of the barriers at the junctions has intermediate values  both the 
thermal effect and the resonant oscillation effect considered in this paper are in effect
simultaneously provided the temperature is near the Thouless temperature. The effects
can be distinguished by decreasing the temperature, which results in a decrease of the
amplitude of the conductance oscillations in the case of the thermal effect (vanishing
at zero temperature) while the resonant amplitude of the conductance increases and has its
largest value at zero temperature.

This work was supported by the Swedish Royal Academy of Sciences (KVA) 
and by the Swedish Natural Science Research Council (NFR).

	\section*{APPENDIX}\label{appendix A}
	\subsection{\label{scattering matrix}Elements of the scattering matrix describing coupling}
We are to describe leads of finite width and the scattering matrix should mix different 
modes. The $S$ matrix will then be of size $3N_{\perp} \times 3N_{\perp}$ and the unitary
condition for the submatrices of size $N_{\perp} \times N_{\perp}$ gives %
\begin{equation}
\hat s_{11} \hat s_{11}^{\dag }=1-2\hat s_{12}\hat s_{12}^{\dag }\quad
\label{uncon1}
\end{equation}
\begin{equation}
\hat s_{22}\hat s_{22}^{\dag }+\hat s_{23}\hat s_{23}^{\dag }=\hat
1-\hat s_{12}\hat s_{12}^{\dag }\quad 
\label{uncon2}
\end{equation}
\begin{equation}
\hat s_{22}\hat s_{23}^{\dag }+\hat s_{23}\hat s_{22}^{\dag }=-\hat
s_{12}\hat s_{12}^{\dag }\quad 
\label{uncon3}
\end{equation}
\begin{equation}
\hat s_{11}s_{12}^{\dag }+\hat s_{12}\left( \hat s_{22}^{\dag }+\hat
s_{23}^{\dag }\right) =0\quad 
\label{uncon4}
\end{equation}
\begin{equation}
\hat s_{12} \hat s_{11}^{\dag }+\left( \hat s_{22}+\hat s_{23}\right) \hat
s_{12}^{\dag }=0\quad 
\label{uncon5}
\end{equation}
Hence we have 5 matrix equations for 8 matrices (as each matrix has an
independent hermitian and antihermitian part) there are 3 undetermined
matrices. We choose them to be $\hat s_{12}$ and the antihermitian part of $\hat s_{22}$.

As it was said above the antihermitian
part is not determined by the unitary conditions for the matrix $\hat S$. In
general case it is of the same order of magnitude as the hermitian part as
they are connected with the Kramers-Kronig relation. Therefore the matrix
elements of $\hat s_{22}\propto \epsilon_r$. An analogous analysis of the
rest of the equations shows the matrix elements of $\hat s_{23}-\hat 1$, $
\hat \rho _{11}$ and $\hat V$ to be of the same order of magnitude.

If $\hat s_{12}$ and the anti-hermitian part $\hat s_{22}^{(A)}$ commutes and other matrices are expressed in terms of $\hat s_{12}$ they may be simultaneously diagonalized. The $N_{\perp}$ eigenvalues of matrices $\hat s_{ij}$ are denoted $\lambda_{ij}$ were indices numbering the eigenvalues are suppressed. Directly from the unitary conditions we get
\begin{eqnarray}
&|\lambda_{11}|=\sqrt{1-2|\lambda_{12}|^2}&\\
&|\lambda_{23}|=\sqrt{1-|\lambda_{12}|^2-|\lambda_{22}|^2}&\\
&\lambda_{11}^*\lambda_{12}+\lambda_{12}^*(\lambda_{22}+\lambda_{23})=0&\\
&|\lambda_{12}|^2+2|\lambda_{22}||\lambda_{23}|\cos{(\phi_{22}-\phi_{23})}=0&
\end{eqnarray}
which gives the requirement $\phi_{22}-\phi_{23}\in[\pi/2, 3\pi/2]$. Now use a
hermitian $\hat s_{12}$ {\it  i.e. } real eigenvalues, $\phi_{12}=n\pi$.
Put $\phi_{23}=0$ in order to agree with the weak coupling limit where no phase gain is expected in passing the reservoir. 
We get with $\lambda_{12}$ and $\lambda_{22}^{(A)}$ as eigenvalues of known
hermitian matrices
\begin{eqnarray}
|\lambda_{11}|&=&\sqrt{1-2|\lambda_{12}|^2}\\
\lambda_{11}^{(A)}&=&\lambda_{22}^{(A)}\\
\lambda_{11}^{(H)}&=&-\sqrt{|\lambda_{11}|^2-(\lambda_{11}^{(A)})^2}\\
\lambda_{11}&=&\lambda_{11}^{(H)}+i\lambda_{11}^{(A)}\\
\lambda_{22}^{(H)}&=& -{\lambda_{11}^{(H)}\over 2}+\sqrt{{(\lambda_{11}^{(H)})^2\over 4}+{|\lambda_{12}|^2\over 2}}\\
\lambda_{22}&=&\lambda_{22}^{(H)}+i\lambda_{22}^{(A)}\\
\lambda_{23}&=&\sqrt{1-|\lambda_{12}|^2-|\lambda_{22}|^2}
\end{eqnarray}
A symmetric random matrix with normally distributed elements will have real eigenvalues distributed according to the semicircle law. We have used values of mean 0 and variance 1 creating random matrices $\hat s_{12}$ and $\hat s_{22}^{(A)}$.
\begin{equation}
s^{(nm)}_{ij}=\langle{\phi_n}|\hat s_{ij}|{\phi_m}\rangle
\end{equation}
where $|{\phi_m}\rangle$ is a complete set of vectors
\begin{equation}
\langle{\phi_m}|=\sum_{\alpha=1}^{N_{\perp}} \gamma_\alpha^{(m)}|{\psi_\alpha}\rangle
\end{equation}
where $|{\psi_\alpha}\rangle$ are eigenvectors to $\hat s_{ij}$. If the corresponding eigenvalues are called $\lambda_\alpha$  
\begin{equation}
s^{(nm)}_{ij}=\sum_{\alpha=1}^{N_{\perp}} \sum_{\beta=1}^{N_{\perp}} \gamma_\alpha^{(m)} \gamma_\beta^{(n)*}\langle{\psi_\beta}|\hat s_{ij}|{\psi_\alpha}\rangle=\sum_{\alpha=1}^{N_{\perp}}  \gamma_\alpha^{(m)} \gamma_\alpha^{(n)*}\lambda_\alpha
\end{equation}
With elements in the matrix $\hat s_{ij}$ of order unity 
\begin{equation}
\sum_{\alpha=1}^{N_{\perp}} |\gamma_\alpha^{(m)}|^2=1
\end{equation}
we get $\gamma_\alpha^{(m)}\propto 1/\sqrt{N_{\perp}}$. Our random matrix $\hat s_{12}$ will be 
multiplied by $\sqrt{\epsilon_r/N_{\perp}}$ before eigenvalues are calculated. Then by setting 
$\tilde\epsilon=\epsilon_r/\epsilon_c=1$ and approaching the strong coupling limit a maximum 
value $\epsilon_c$ is found fulfilling the unitary conditions. The strength of the coupling of 
the reservoirs is now parametrized by $\tilde\epsilon\in[0,1]$. The matrix $\hat s_{22}$ is in 
the weak coupling limit for one channel seen to be proportional to $\epsilon_r$ \cite{Buttiker84} 
and therefore the random matrix giving $\lambda_{22}^{(A)}$ is multiplied by 
$\epsilon_r/\sqrt{N_{\perp}}$.
Then by using the eigenvectors of the matrix $\hat s_{12}$ we transform all the
 matrices $\hat s_{ij}$ back to the initial representation in which they are not
diagonal and their matrix elements are the probability amplitudes of
scattering to the respective transverse modes. 

To realize another type of scattering matrix to describe the coupling to the reservoirs we do as 
follows.
The essential features of the junction are coupling to electron reservoirs and mixing between 
modes, both features may be parametrized by the strength of the coupling $\tilde\epsilon$. The
 coupling to the reservoirs is described by B\"uttiker matrices.
 \cite{Buttiker84} If this is done 
mode by mode there will be no mixing. An additional unitary matrix is used to mix modes. This 
matrix has all diagonal elements equal to each other and all off-diagonal elements equal to each 
other, describing scattering into the same mode and mixing between modes respectively. By keeping 
the elements equal an isotropic situation is simulated where scattering into any mode is possible. 
The elements $u$ of the $N_{\perp}\times N_{\perp}$ unitary matrix must fulfill
\begin{equation}\label{UnitaryU}
|u_{ii}|^2+(N_{\perp}-1)|u_{ij}|^2=1
\end{equation}
\begin{equation}
u_{ii}^*u_{ij}+u_{ij}^*u_{ii}+(N_{\perp}-2)|u_{ij}|^2=0
\end{equation}
This gives $|u_{ij}|^2 \leq 1/(N_{\perp}-1)$, the phases of $u_{ij}$ and $u_{ii}$ are connected by
\begin{equation}\label{unitary.angle}
\varphi_{ij}=\varphi_{ii}-\arccos\left(-\frac{(N_{\perp}-2)}{2}\frac{|u_{ij}|}{|u_{ii}|}\right)\\
\end{equation}
we note that for $N_{\perp}>4$, $|u_{ij}|$ and $|u_{ii}|$ are not allowed to have the same value. 
We wish to consider large $N_{\perp}$ and write the elements
\begin{eqnarray}
{|u_{ii}|}^2&=&1-c\tilde\epsilon\\
{|u_{ij}|}^2&=&\frac{c\tilde\epsilon}{N_{\perp}-1}
\end{eqnarray}
in order to agree with the limit $\tilde\epsilon=0$ when mixing is expected to be absent since the 
waves in the decoupled sample do not feel the reservoirs. The condition Eq.~({UnitaryU}) gives
\begin{equation}
c=\frac{4(N_{\perp}-1)}{N_{\perp}^2}
\end{equation}
The parameters are the coupling $\tilde\epsilon$, the number of modes $N_{\perp}$ and the angle 
$\varphi_{ii}$. We write the angle $\varphi_{ii}\in\tilde\epsilon[0, 2\pi]$ to agree with the weak 
coupling limit where no phase gain is expected. By using different angles we get an ensemble of 
matrices describing different samples.

The eigenvalues of the unitary $\hat U$ all have amplitudes of length unity. This means that all 
modes will be open for transmission. \cite{Imry86} Opening of transmission channels has been 
observed in experiments. \cite{Mur96}

To describe coupling the results by B\"uttiker \cite{Buttiker84} are used in diagonal matrices. 
All these matrices $\hat s_{ij}$ are multiplied by $\hat U$. 

\subsection{\label{barrier}A barrier at the NS-interface}
A Schottky barrier\cite{Blonder82} at the interface of a semiconductor and a metal can be 
characterized with a scattering matrix
\begin{equation}
\hat s^{(0)}=-\left(
\begin{array}{cc}
r_0 & t_0 \\
-t_0^{*} & r_0^{*}
\end{array}
\right) ,\,\,\,\,\,\,|r_0|^2+|t_0|^2=1
\label{norm}
\end{equation}
that connects the constant factors $a_1$ and $b_1$ of the plane waves coming
in and going out of the barrier inside the semiconductor, respectively, with
those $a_2$ and $b_2$ inside the metal:
\begin{equation}
b_i=\sum_{k=1}^2s_{ik}^{(0)}a_k,\,\,\,\,\,\,\ i=1,2
\label{norm2}
\end{equation}
(Hence $|t_0|^2$ is the transparency of the barrier) When the metal is in
the superconducting state the matrix of reflection of the semiconductor
charge carriers at the N-S boundary (the semiconductor is on the right and
the superconductor is on the left of the N-S boundary) is
\begin{equation}
\hat \eta =e^{i\Psi }\left(
\begin{array}{ll}
r_N & r_A \\
-r_A^{*} & r_N^{*}
\end{array}
\right) 
\label{nandr}
\end{equation}
where
\begin{equation}
e^{i\Psi}=i\frac{
\sqrt{
|t_0|^4+4|r_0|^2\sin^2\psi_{E}
}}{
e^{i\psi_{E}}-|r_0|^2e^{-i\psi_{E}}
}\label{nandr1}
\end{equation}
\begin{equation}
r_N=r_0^{*}\frac{2\sin\psi_{E}}{\sqrt{|t_0|^4+4|r_0|^2\sin ^2\psi_{E}}}
\label{nandr2}
\end{equation}
\begin{equation}
r_A=it_0^2\frac{e^{i\phi }}{\sqrt{|t_0|^4+4|r_0|^2\sin ^2\psi _{E}}}
\label{nandr3}
\end{equation}
\begin{equation}
e^{i\psi _{E}}=\frac{|\Delta |}{E-i\sqrt{|\Delta |^2-E^2}}
\label{nandr4}
\end{equation}
Here $|\Delta |$ and $\phi$ are the modulus and the phase of the
superconducting gap, respectively, $E$ is the electron energy measured from
the Fermi level. From Eq.~(\ref{nandr}-\ref{nandr4}) it is straightforward to
see that $|r_A|^2+|r_N|^2=1$.
	\subsection{\label{active channels}Active channels}
As for $N_{\perp}\gg 1$ the set of equations Eq.~(\ref{major}) can not be analytically solved 
and the amplitudes $\gamma _n$ can
not be explicitly found we estimate the number $N_R$ of transverse modes
inside the resonant region $(-\epsilon_r\hbar v_F/L,\epsilon_r\hbar
v_F/L)$ and use Eq.~(\ref{G}) to get the conductance to within a factor of
the order of unity. We determine $N_R$ in the following way.
\begin{equation}
N_R=\int\limits_{-\infty }^\infty dE\frac{\epsilon_r^2}{(E L/\hbar
v_F)^2+\epsilon_r^2}\nu (E)  
\label{resnumb}
\end{equation}
where $\nu (E)$ is the state density function
\begin{equation}
\nu (E)=\sum\limits_l \sum\limits_{n=1}^{N_{\perp}}\delta
(E-E_{n,l})=\sum\limits_{n=1}^{N_{\perp}}|\frac{\partial Q_n}{\partial E}|\delta (Q_n);
\label{apnu}
\end{equation}
Here the spectrum function $Q_n$ is determined by Eq.~(\ref{sfg}).

To find the state density function $\nu (E)$ we use the method developed in
Ref.~\onlinecite{slutskin}. As $\frac{\partial \varphi _{-}}{\partial E}=\pm (\hbar
v)^{-1}L$ with $v=\hbar\sqrt{k_F^2-k_{\perp }(n)^2}/m$ the factor $\partial Q_n/\partial E$ in 
(\ref{apnu}) is
a trigonometrical function of ${\pm }\varphi _{\pm }$ as well as $Q_n$ is, and
it is productive to expand $\nu $ into Fourier series in $\varphi _{\pm }$
and write it as follows.
\begin{equation}
\nu (E)=\sum_{n=0}^{N_{\perp}}\sum_{s=-\infty }^\infty \sum_{k=-\infty
}^\infty A_{s,k}e^{i(s\varphi _{-}+k\varphi _{+})}\quad 
\label{apnuFur}
\end{equation}
\begin{equation}
A_{s,k}=(2\pi)^{-2}\int_0^{2\pi }d{\bar\varphi}_{+}\int_0^{2\pi }d{\bar\varphi} _{-}|
\frac{\partial Q_n}{\partial E}|\delta (Q_n)e^{-i(s{\bar\varphi}_{-}+k{\bar\varphi}
_{+})}\quad 
\label{apamp}
\end{equation}
In this paper we assume the length $L$ and the width $d$ of the sample to
satisfy the inequality
\begin{equation}
\frac dL\gg \sqrt{\frac{\lambda _F}L}  
\label{inequality1}
\end{equation}
Using this inequality, Eq.~(\ref{fastosc}) and the estimation of Eq.~(\ref{sfin}) in 
Appendix~D one sees that the main contribution to the state density 
function Eq.~(\ref{apnu}) is of the terms
\begin{eqnarray}
\label{apnumaj}
A_{s,0}&=&
\frac L{(2\pi )^2\hbar v}
\int_0^{2\pi }d{\bar\varphi}_+
\int_0^{2\pi}d{\bar\varphi}_-
|\sin {\bar\varphi}_-|e^{-is{\bar\varphi}_-}\\
&&\times\delta (
\cos {\bar\varphi}_-
-|{ r_{N1}}||{ r_{N2}}|\cos {\bar\varphi}_+
+|{ r_{A1}}||{ r_{A2}}|\cos \phi)
\nonumber
\end{eqnarray}
Performing integration with respect to ${\bar\varphi}_{-}$ in (\ref{apnumaj}) and
over $E$ in (\ref{resnumb}) with application of (\ref{apnuFur}), (\ref{apamp}) one
obtains the conductance
\begin{equation}
G=N_{\perp}\frac{e^2}{2\pi \hbar }\epsilon_r\sum_{s=-\infty }^\infty
e^{-2|s|\epsilon_r}\int_0^{2\pi}\cos s\varphi _1(\varphi
_{+})d\varphi _{+}  
\label{22}
\end{equation}
\[
\varphi _1(\varphi _{+})=\arccos \Bigl(|{ r_N}_1||{ r_N}_2|\cos
(\varphi _{+})-|{ r_A}_1||{ r_A}_2|\cos \phi \Bigr)
\]
if $0\leq \varphi_{-}\leq \pi $, and
\[
\varphi _1(\varphi _{+})=2\pi -\arccos \Bigl(|{ r_N}_1||{ r_N}_2|\cos
(\varphi _{+})-|{ r_A}_1||{ r_A}_2|\cos \phi \Bigr)
\]
if $\pi \leq \varphi _{-}\leq 2\pi $.
Performing summation in (\ref{22}) one gets
\begin{eqnarray}
\label{Gint}
G&=&N_{\perp}\frac{e^2}{\pi \hbar }\epsilon_r \times\\
&&\int_0^{2\pi }\frac{\epsilon_r d\varphi_+}{1-|{
r_N}_1||{ r_N}_2|\cos \varphi _{+}+|{ r_A}_1||{ r_A} _2|\cos \phi +\epsilon_r^2/2}
\nonumber
\end{eqnarray}
as $|{ r_N}_{1,2}|\leq 1$ integration in (\ref{Gint})
gives
\begin{equation}
G=N_{\perp}\frac {e^2}{\pi\hbar}\frac{\epsilon_r^2}{\sqrt{(1+|{ r_A}_1||
{ r_A}_2|\cos \phi +\epsilon_r^2/2)^2-|{ r_N}_1|^2|{ r_N}_2|^2)}}
\label{Gapp}
\end{equation}
If the boundaries are symmetric, that is ${ r_{N1}=r_{N2}}$, the
conductance is
\begin{equation}
G=N_{\perp}\frac {e^2}{\pi\hbar}\frac{\epsilon_r^2}{\sqrt{(1+|{ r_A}%
|^2\cos \phi +\epsilon_r^2/2)^2-|{ r_N}|^4)}}
\end{equation}

\subsection{\label{fast.oscillation} Fast oscillating terms}
In this Appendix we evaluate a sum of fast oscillating terms
\begin{equation}
S=\frac 1{N_{\perp}}\sum\limits_{n=0}^{N_{\perp}}e^{i L
[k_F^2-(\frac \hbar {d}n)^2]^{1/2}}  \label{fastosc}
\end{equation}
that appears in the density state function Eq.~(\ref{apnu}) and the
transition probability Eq.~(\ref{gamosc}), Eq.~(\ref{trnp}). Using the Poisson
formula one can write
\begin{equation}
S=\lim\limits_{\delta \rightarrow 0}\frac 1{N_{\perp}}\sum\limits_{k=-\infty }^\infty \ \int\limits_0^{N_{\perp}}dxe^{i\lambda
\sqrt{1-x^2/\alpha ^2}+i2\pi kx}e^{-\pi |k|\delta }  \label{poisson}
\end{equation}
Here
\begin{equation}
\lambda =Lk_F ,\,\,\,\alpha =k_Fd/\pi   \label{lambda}
\end{equation}
As $\lambda \gg 1$ one can use the saddle point method to get
\begin{eqnarray}
\label{sadle}
S&=&\frac 1{N_{\perp}}\frac \alpha {\sqrt{\lambda }}\sqrt{\pi }e^{-i\pi
/4}\times \\
&&\sum\limits_{k=-\infty }^\infty \left( \frac{\lambda ^2}{\lambda ^2+(2\pi
\alpha k)^2}\right) ^{3/4}e^{i\sqrt{\lambda ^2+(2\pi \alpha k)^2}}
\end{eqnarray}
From Eq.~(\ref{sadle}) it follows that terms with $k$ which are less or of
the order of $\lambda /\alpha $ contribute to the sum and, therefore, $S$ is
less than $\sqrt{\lambda }/$ $N_{\perp}$. As $N_{\perp}\sim k_Fd$ we
have the following estimation of the sum of fast oscillating terms.
\begin{equation}
S\leq \sqrt{Lk_F}/\left( k_Fd\right)  \label{sfin}
\end{equation}
Therefore the sum of fast oscillating terms can be neglected if
\begin{equation}
\frac dL\gg \sqrt{\frac{\lambda _F}L}  \label{inequality2}
\end{equation}


\begin{thebibliography}{30}
\bibitem{fot1}
Recent experiments \cite{Petrashov97,Hideaki97}
show a distinct shift in the phase dependence from 
what is expected for the thermal effect (the conductance maxima 
appear for even multiples of $\pi$) to the pattern typical for the giant 
oscillation effect (the conductance maxima are at odd multiples of $\pi$) 
when the temperature is lowered.
It is worth noticing that in cases where no Schottky barrier is expected
at the junctions between sample and reservoirs some scattering is stillö
caused by the junction geometry.
In this situation  backscattering 
is possible only from a region within a distance of order $\sim \lambda_B$
from the junction edges. Under this condition the zero-temperature 
amplitude is not giant in the sense used in this introduction, but it is
nevertheless large.
\bibitem{note_order} 
Although in a given experiment the matrix elements are fixed and uniqely 
determined, they are sample dependent.
We keep in mind that the conductance which
we calculate should be ensemble averaged with respect to such uncontrollable
variations. If in this procedure one formally considers this matrix to 
be random and its elements
distributed around $1/\sqrt{N}$ to satisfy the unitary conditions
(\ref{uncon1})-(\ref{uncon5}), then its eigenvalues will be distributed in the interval
[-1, 1] according to the semicircle law \cite{semicirclelaw}. Hence, in such an averaging
procedure the most representative matrices would have eigenvalues of order unity.
\bibitem{fot2}
We assume that $E_{c}\ll |\Delta| $ so that Andreev reflection is prominent 
on the energy scale given by $E_c$.
Physically this  means that the distance between the superconductors is much
larger than the penetration length into the normal metal of the superconducting correlations.
In experiments this condition is easily satisfied since $E_c/k_B \sim$~0.1~K.
\bibitem{fot3}
According to Lambert \cite{Lambert1}, when a voltage is applied between the
reservoirs the chemical potential of the superconductors  (the chemical potentials of the
semiconductors and  the normal part of the sample are equal) must be  chosen self-consistently with
the condition of conservation of the current outgoing from and incoming to the reservoirs via the
sample. The superconductor chemical potential as a result lies between the chemical potentials of
the reservoirs. Hence electrons are emitted from the reservoir with the higher chemical potential,
and holes from the  reservoir with the lower potential.
\bibitem{fot4}
It interesting to note that this peak splitting
takes place also for a ballistic sample. However, in this case, in contrast to the
diffusive sample, the contribution to the conductance due to transmission of an electron to
the second reservoir via the ballistic sample is not suppressed. As a result, in the general
case for the ballistic system the conductance is not equal to zero at $\phi=\pi$. However
the value of the conductance at $\phi=\pi$ is controlled by the parameter $k_F l_3$
($l_3$ is the distance between the junctions) and is equal to zero when $k_F l_3=\pi/2$
(modulo $\pi$) [see the discussion at the end of Section \ref{coupling}).
\bibitem{fot5}
This means that for the $N_{\perp}\gg 1$
transverse modes  a summation with  respect to starting points of the  paths
is equivalent to an impurity averaging of the conductance.
This is a manifestation of the fact that the conductance is a self-averaging 
quantity.
\end{thebibliography}
\end{document}